\pdfoutput=1 
\documentclass[graybox,envcountchap,sectrefs]{svmono}

\usepackage{mathptmx}
\usepackage{helvet}
\usepackage{courier}
\usepackage{type1cm}

\usepackage{makeidx}         
\usepackage{graphicx}        
\usepackage{multicol}        
\usepackage[bottom]{footmisc}

\usepackage{amssymb, amsmath, amsfonts, amsopn} 
\usepackage{natbib}
\usepackage{deluxetable}
\usepackage{hyperref}
\usepackage{rotating}					
\usepackage{epsfig}

\usepackage{color}
\definecolor{Blue}{rgb}{0.3,0.3,0.9}

\usepackage{pdflscape}
\usepackage{longtable}

\usepackage[nottoc]{tocbibind}
\makeindex             

\newcommand{\teff}{$T_{\rm eff}$}			

\newcommand{\rsun}{R$_{\odot}$}			
\newcommand{\lsun}{L$_{\odot}$}			
\newcommand{\rearth}{R$_{\rm Earth}$}		
\newcommand{\mearth}{M$_{\rm Earth}$}		

\begin{document}

\author{Kaspar von Braun \& Tabetha S. Boyajian}

\title{Extrasolar Planets and Their Host Stars}

\maketitle

\frontmatter

%
%

\preface

In astronomy or indeed any collaborative environment, it pays to figure out with whom one can work well. From existing projects or simply conversations, research ideas appear, are developed, take shape, sometimes take a detour into some unexpected directions, often need to be refocused, are sometimes divided up and/or distributed among collaborators, and are (hopefully) published. After a number of these cycles repeat, something bigger may be born, all of which one then tries to simultaneously fit into one's head for what feels like a challenging amount of time. That was certainly the case a long time ago when writing a PhD dissertation. 
\\
\\
Since then, there have been postdoctoral fellowships and appointments, permanent and adjunct positions, and former, current, and future collaborators. And yet, conversations spawn research ideas, which take many different turns and may divide up into a multitude of approaches or related or perhaps unrelated subjects. Again, one had better figure out with whom one likes to work. And again, in the process of writing this Brief, one needs create something bigger by focusing the relevant pieces of work into one (hopefully) coherent manuscript. It is an honor, a privilege, an amazing experience, and simply a lot of fun to be and have been working with all the people who have had an influence on our work and thereby on this book. 
\\
\\
To quote the late and great Jim Croce: "If you dig it, do it. If you really dig it, do it twice."





\vspace{\baselineskip}
\begin{flushright}\noindent
Pasadena \& Baton Rouge,\hfill {\it Kaspar von Braun}\\
July 2017\hfill {\it Tabetha Boyajian}\\
\end{flushright}

%
%

\extrachap{Acknowledgements}

First and foremost, we would like to express our sincere gratitude to our principal collaborators, Gerard T. van Belle, Stephen R. Kane, Gail Schaefer, Andrew Mann, Gregory Feiden, David R. Ciardi, Tim White, and Theo ten Brummelaar, for their scientific contributions to this work over the course of the past seven years. We furthermore thank the CHARA gang (Chris Farrington, PJ Goldfinger, Nic Scott, Norm Vargas, Olli Majoinen, Judit Sturmann, Laszlo Sturmann, Nils Turner) for their tireless and invaluable support of observing operations at the Array. Thanks to Barbara Rojas-Ayala, Phil Muirhead, Eric Gaidos, Daniel Huber, Hal McAlister, Stephen Ridgway, Sean Raymond, Douglas Gies, Orlagh Creevey, and Lisa Kaltenegger for multiple insightful and useful discussions on various aspects of this work. 
This work is based upon observations obtained with the
Georgia State University Center for High Angular
Resolution Astronomy Array at Mount Wilson Observatory.
The CHARA Array is supported by the National Science
Foundation under Grant No. AST-1211929.
Institutional support has been provided from the GSU
College of Arts and Sciences and the GSU Office of the
Vice President for Research and Economic Development.
This research made use of the SIMBAD and VIZIER Astronomical Databases, operated at CDS, Strasbourg, France (http://cdsweb.u-strasbg.fr/), and of NASA's Astrophysics Data System, of the Jean-Marie Mariotti Center \texttt{SearchCal} service (http://www.jmmc.fr/searchcal), co-developed by FIZEAU and LAOG/IPAG. This publication makes use of data products from the Two Micron All Sky Survey, which is a joint project of the University of Massachusetts and the Infrared Processing and Analysis Center/California Institute of Technology, funded by the National Aeronautics and Space Administration and the National Science Foundation. This research made use of the NASA Exoplanet Archive \citep{ake13}, which is operated by the California Institute of Technology, under contract with the National Aeronautics and Space Administration under the Exoplanet Exploration Program. This work furthermore made use of the Habitable Zone Gallery at hzgallery.org \citep{kan12}, the Exoplanet Orbit Database and the Exoplanet Data Explorer at exoplanets.org \citep{wri11b}, and the Exoplanet Encyclopedia at exoplanet.eu \citep{sch11}.


\tableofcontents

\mainmatter
%
%
%
\chapter{Introduction}
\label{chapter:introduction} 




From the dawn of astronomy, planets around other stars have captured the public attention because they are something everyone can imagine -- they are a place rather than merely a concept. From science fiction, we can picture ourselves taking a step out of a spaceship and looking around on another planet. Long gone are the days when every new discovery of one such planet outside our solar system would make the news. There are now thousands of them, either confirmed or validated, and orders of magnitude more candidates. In fact, there is increasing evidence that there are many more planets in the Milky Way than stars (i.e., $>$ hundreds of billions), particularly in orbit around the significantly more abundant, and arguably much more interesting, low-mass stars. 

Without a doubt, all of the gathered insights on exoplanets represent a testament to the tireless work performed by exoplanet pioneers in the astronomical community, fueled by the interest and fascination of the general public in and about extrasolar planets. We now know of planetary system architectures that rival the inventiveness of science fiction material. The large and growing number of studied exoplanets is gradually permitting insight into a variety of statistical and empirical relations such as interdependence of planetary and stellar parameters, though, as always, ``more data are needed." 

The study of every individual planetary system is and will remain a worthwile endeavor, particularly for the more scientifically interesting systems -- however that may be defined. For instance, transiting planets offer more astrophysical insights than non-transiting planets due to the fact that the ratio of planetary to stellar radius and the system's inclination angle with respect to the line of sight are known. In addition, planets around smaller stars tend to be more abundant than planets around larger and hotter stars, and they are smaller themselves, i.e., more likely to be rocky and thus more Earth-like. Thus, they generally represent higher-priority targets of planet surveys since any kind of life can presumably exist on rocky planets but not as easily on gas giants.
Finally, a stellar habitable zone (HZ) is loosely defined here as the range of distances from its respective parent star at which a rocky planet with some atmosphere could host liquid water on its surface. This obviously makes planets in their respective HZs particularly interesting. 

The large and increasing number of detected planets presents the challenge and opportunity of looking at the situation statistically, provided one does not introduce a bias into one's experiment. Astronomers are very good at correlating everything with everything and seeing what happens. One approach along those lines is to examine planetary masses or radii since they depend upon their stellar counterparts. Physical parameters of planets are almost always direct functions of physical parameters of their host star. For instance, the size of a transiting planet is a function of the observed transit depth, the size of the parent star, and the ratio of surface temperatures of planet and star. The mass of a planet is a function of the observed radial velocity amplitude, the stellar mass, and the (often unknown) inclination angle of the planetary orbit with respect to the line of sight. In addition to such direct dependencies, the star is frequently the only visible component in the system, the dominant source of energy, and it contains most of the system's mass. Consequently, parent stars heavily influence every aspect of planetary physics and astrophysics. 







These parent stars exist in very many different flavors: big, small, hot, cool, young, old, with various chemical compositions, as parts of multiple star systems (hierarchically bound by gravity) or as single stars, located within the Galactic disk or halo, as members of a star cluster (open or globular) or field stars, etc. Planets also show a large variety of physical and chemical properties. Some are, but many are not, akin to the Solar System planets we know. Some are smaller than Mercury, some are larger than Jupiter, some have surface temperatures of many thousands of degrees, some have icy surfaces, some don't have surfaces at all. Some have thick atmospheres, some have extremely strong jets and winds, some have atmospheres that are evaporating, some may not have any atmospheres. Some may have water clouds in their atmospheres, some may have liquid water on their surfaces. 



Knowledge of the radii and effective temperatures of the parent stars is of considerable importance as these parameters define the radiation environment in which the planets reside. Thus, they are required for the calculation of the circumstellar habitable zone's (HZ) location and boundaries as well as the planetary equilibrium temperatures. Furthermore, the radii and densities of any transiting exoplanets, which provide profound insights into planet properties, such as their interior structures and bulk densities, are direct functions of the radius and mass of the respective parent star. 

{\it In order to understand the planet, you will need to understand its parent star.} This sentence shall serve as the motivation for the work presented here. Our goal is to characterize exoplanet host stars to provide insights into the properties of the planets themselves. In the process, we try to rely on as few as assumptions about the stars as possible, i.e., to be as direct as possible. 

We focus on the  determination and calculation of stellar parameters in Chapter \ref{chapter:parameters}, in particular the empirical measurements of stellar radii and effective temperatures via interferometry and stellar energy distribution fitting. The same chapter describes how directly determined stellar parameters are used to characterize various aspects of exoplanetary systems, e.g., how an exoplanetary system's habitable zone is dependent on stellar luminosity and effective temperature. To give insight into the status of the field of stellar characterization, we provide a table of currently known stars with directly determined radii in Chapter \ref{chapter:systems} where we also present details and/or background on a number of individual exoplanetary systems. Chapter \ref{chapter:future} briefly outlines future work, such as current high-priority targets, optimization of methods, efforts to reduce unknown systematics, application of our insights to stars that are too faint or too distant to be studied empirically. We briefly summarize and conclude in Chapter \ref{chapter:conclusion}.

\chapter{The Determination of Stellar and Planetary Astrophysical Parameters}
\label{chapter:parameters} 



{\it In order to understand the exoplanet, you need to understand its parent star.} 
\\
\\
Stellar characterization is an old and established field within astronomy -- which is not to say that it is easy or that stars are sufficiently well understood in general. Many determinations of stellar parameters rely, by necessity, on indirect measurements and/or assumptions, especially if direct data are sparse, such as is the case for low-mass stars, by which we mean here stars whose outer layers are fully convective, making energy transport to the stellar surface poorly understood. Calibrations linking non-observable quantities to directly determined parameters have to constantly be revised and improved. Constraints imposed by new data on the hugely important stellar models serve to increase their predictive power. In this Chapter, we briefly describe our methods used for the direct determination of stellar astrophysical parameters, elaborate on their usefulness, and discuss the applicability of the parameters to other insights into exoplanet science.  


\section{Stellar Radius}
\label{chapter:parameters:sec:radius}


Stellar radii can be measured directly only by few methods, among which the most widely used ones are (1) the study of eclipsing binary stars, (2) asteroseismology, and (3) very-high angular resolution measurements, mostly performed via interferometry. 

The use of eclipsing binaries (EBs) to study stars and stellar systems is one of the oldest methods of studying stellar parameters in the field of astronomy. It is based on the study of the system light curve in which the eclipse durations and depths provide insight into the relative sizes of the components. There are many excellent publications and reviews of this appraoch to which we refer the reader for more information, such as \cite{and91} and \cite{tor10}. 

The study of asteroseismology is based on the analysis of stellar oscillations and measures stellar bulk density and variations of the sound speed through the interior of the star. These insights can actually probe stellar chemical composition, structure, core conditions and evolutionary age, and thus, ultimately stellar mass and age. High-precision photometric surveys (Kepler, CoRoT, TESS, PLATO, CHEOPS, etc) looking for exoplanet transits automatically provide or will provide data that can be used to perform asteroseismology, making this technique quite powerful. We refer the reader to the very good reviews by \citet{hub13} and \citet{hub16} and references therein for more detailed information.  

For the purposes of this publication, we will focus on the method of interferometry to directly determine the angular radii of stars, the results of which are then convolved with empirically determined distances and bolometric flux measurements to calculate stellar physical radii, effective temperatures, and luminosities (\S \ref{chapter:parameters:sec:temp}). 

\begin{figure}
\includegraphics[width=\linewidth]{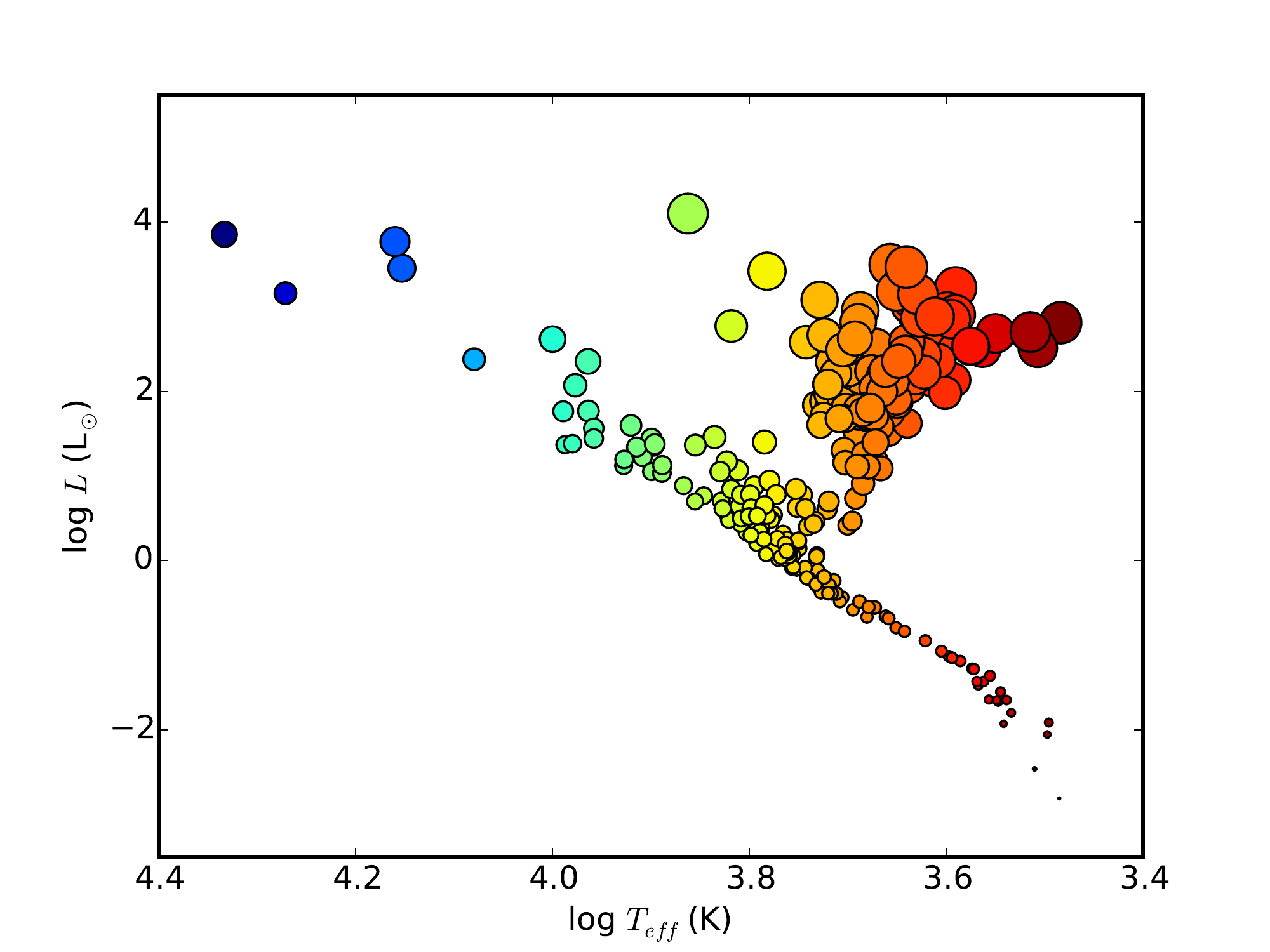}
\caption{Empirical H-R Diagram for all stars with interferometrically determined stellar radii with random uncertainties smaller than 5\%, stellar radii $<$ 100 \rsun, and distances up to 150 pc (status 1 Nov 2016). The diameter of each data point is representative of the logarithm of the corresponding stellar radius. Error bars in effective temperature and luminosity are smaller than the size of the data points. The color of each data point is representative of the respective stellar effective temperature. 
For more details, also on the references for stellar angular diameters, effective temperatures, and luminosities, see \S \ref{chapter:systems:sec:status}, Table \ref{tab:systems:dwarfs}, and Fig. \ref{fig:HRD_EHS_blue}.\label{fig:HRD_allstars}}
\end{figure}

The first stellar diameter determined by interferometry was that of the red giant star Betelgeuse by using a rigid, 20-feet long beam as the baseline and the 2.5m Telescope on Mount Wilson \citep{mic21,mic21a}. Motivated by the success of radio-wavelength interferometry in the 1950s, the first stellar diameters measured at visible wavelengths with two separate telescopes were published in \citet{bro74} and later \citet{lab75,lab78}. Due to the relative infancy of the field at that time, the interferometers were limited to bright stars ($B<2.5$), essentially all of which were giant stars. Since then, the majority of the diameters of {\it main-sequence} stars that were determined interferometrically were measured with the Georgia State University's CHARA Array\footnote{Center for High Angular Resolution Astronomy; see \citet{ten05,ten13c} for description of the array and instruments.}. We show in Figure \ref{fig:HRD_allstars} the current status (1 Nov 2016; based on data from Table \ref{tab:systems:dwarfs}) of stars with high-precision interferometrically determined diameters
with random uncertainties smaller than 5\%, stellar radii $<$ 100 \rsun, and distances up to 150 pc. The diameter of each data point is indicative of the logarithm of the respective stellar diameter, ranging from around 0.15 \rsun  through 100 \rsun. Error bars in this diagram are smaller than the size of the data points. The color of each data point is representative of the respective stellar effective temperature.

The concept of measuring a stellar diameter using interferometry dates back to the 19th century, when Hippolyte Fizeau proposed performing Young's double-slit experiment in front of a primary mirror of a telescope, which was performed by Edouard Stephan in 1874 \citep{len14}. The study of stellar angular diameters via interferometry is based on determining the coherence of light from a star at two or more points separated by one or more baselines. Since all telescopic mirrors have a finite area, it is fair to say that all telescopes are actually interferometers. For the sake of this qualitative discussion and for the purpose of this publication, however, we will assume the case of two separate telescopes separated in distance by a single baseline.

\begin{figure}
\includegraphics[width=\linewidth]{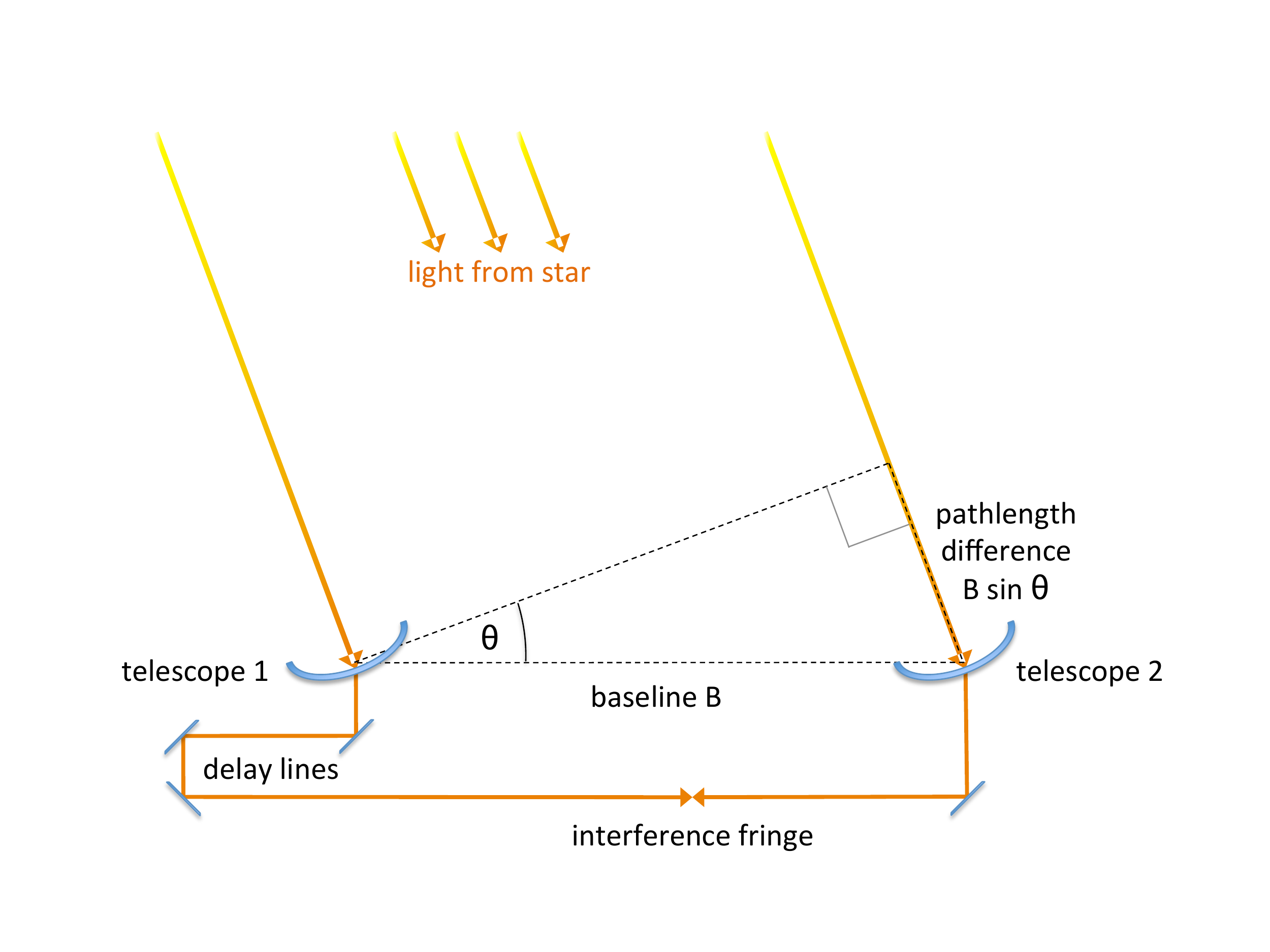}
\caption{Schematics of a 2-telescope interferometric array. The travelled distance of the star light is larger at telescope 2 by the path length difference of B $\sin \theta$. This increased path length is compensated by the use of variable-length delay lines between telescope 1 and the position of the interference fringe (Fig. \ref{fig:fringe}) at which light is coherent. Note that in reality, all telescopes in an interferometric array have delay lines in their optical paths since otherwise, only very reduced parts of the sky would be accessible for observations. Note also that the amount of delay that can be added governs the observable range of zenith distances. \label{fig:schematics}}
\end{figure}

\begin{figure}
\includegraphics[width=\linewidth]{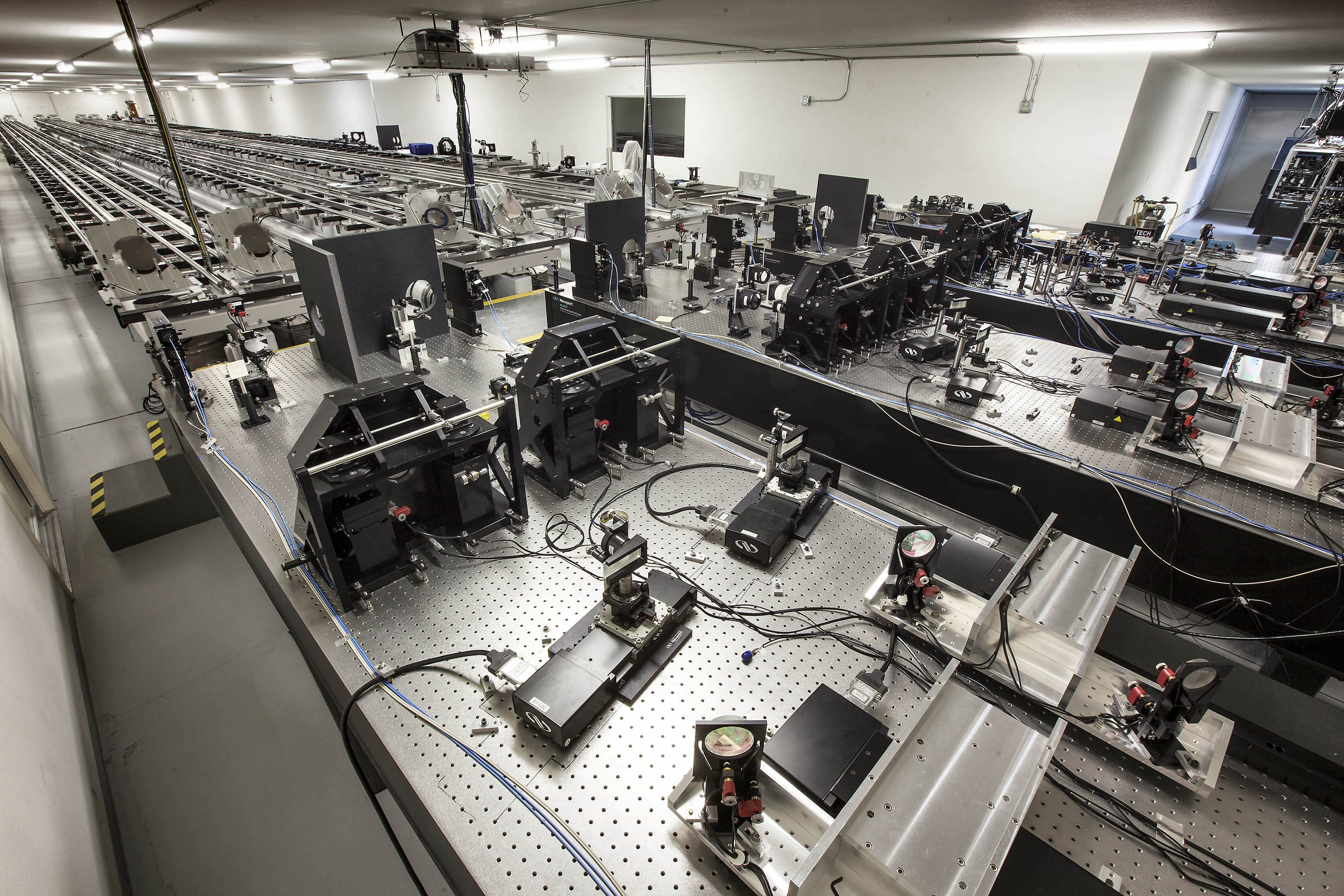}
\caption{Picture of optical delay line setup at the CHARA Interferometric Array. The delay lines extend toward the top left of the picture from the optical benches in the bottom right. Mirrors/lenses on the optical benches direct the light path toward the beam combiners and detectors, barely visible in the picture at the top right. Photo courtesy of Steve Golden. \label{fig:ople1}}
\end{figure}

\begin{figure}
\includegraphics[width=\linewidth]{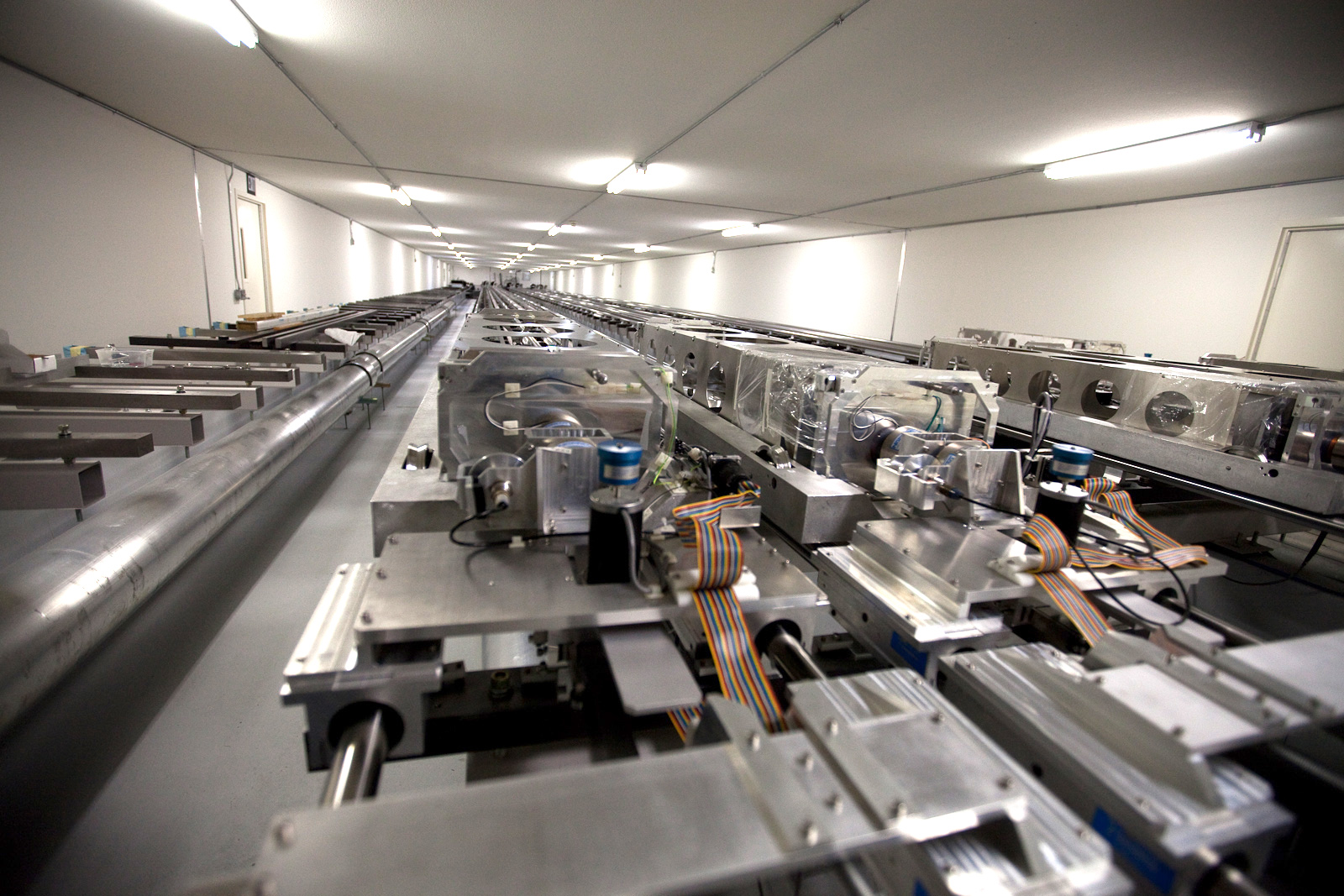}
\caption{Picture of optical delay lines at the CHARA Interferometric Array as seen from the optical benches shown in Figure \ref{fig:ople1}. The silver boxcars with circular cavities on their sides (right side of picture) are the carts that contain mirrors to change the optical path length of the beam. They drive back and forth on tracks and contain a number of metrology mechanisms to adjust path length to a fraction of the operational wavelength. Photo courtesy of Theo ten Brummelaar. \label{fig:ople2}}
\end{figure}

\begin{figure}
\includegraphics[width=\linewidth]{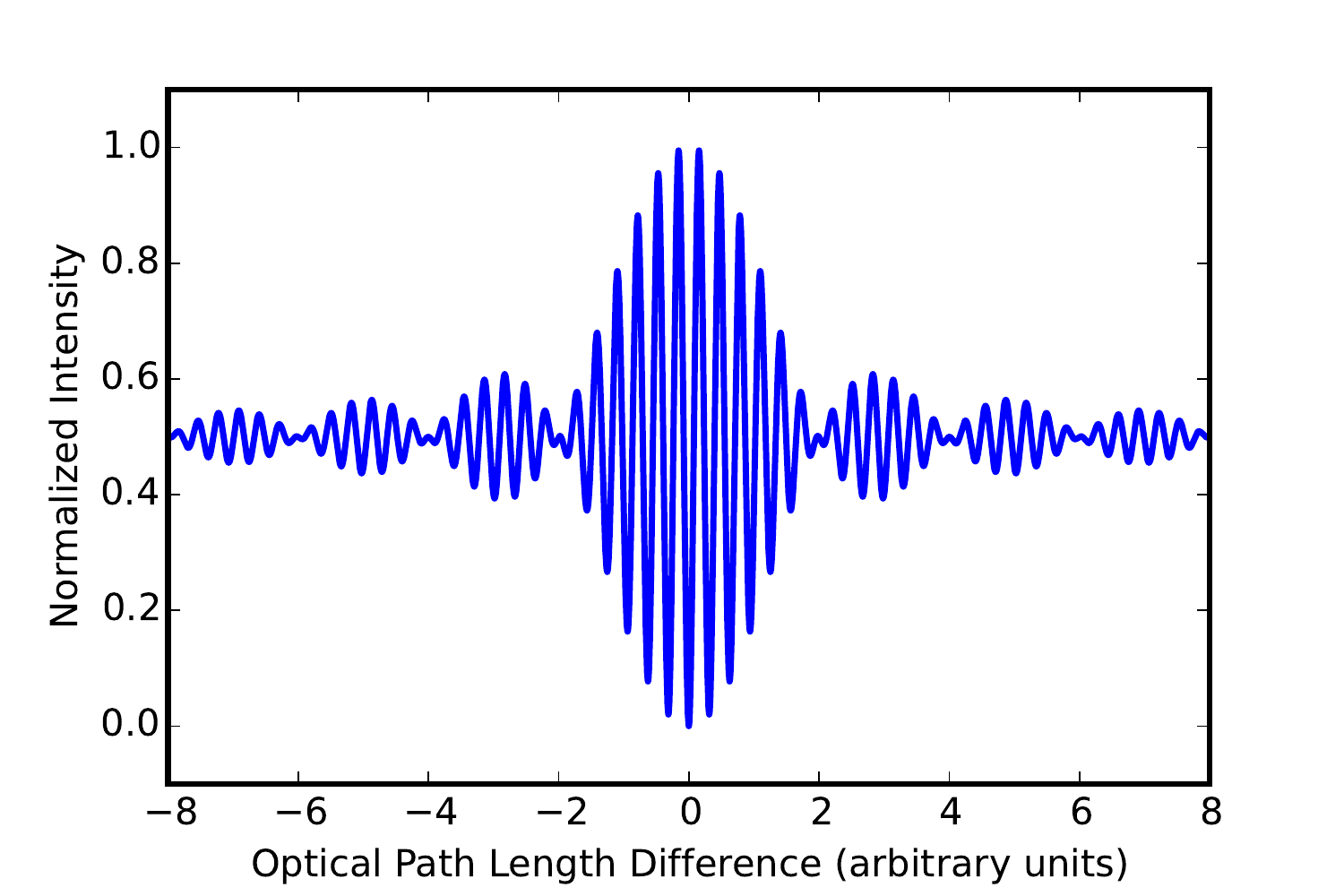}
\caption{Example of an interferometric fringe -- visualizing the dependence of the intensity of the combined light as a function of optical path length difference between two telescopes for a polychromatic source such as a star.\label{fig:fringe}}
\end{figure}

In order for interferometric interferences to occur on the detector where the signals from the two telescopes are combined, the light from the star has to be coherent, that is, all photons have to have travelled exactly -- to a fraction of the operational wavelength -- the same distance between their respective origins on the stellar surface and the detector via the optical paths of the both telescopes (see Figure \ref{fig:schematics}).  This necessitates the use of optical delay lines with which additional path length can be introduced (see Figure \ref{fig:schematics} for a schematic setup and Figures \ref{fig:ople1} and \ref{fig:ople2} for photos of the CHARA optical path length delay facility), which increases with increasing zenith distance and longer baselines. 
It also requires exquisite engineering precision in the optical quality of the telescopes and the entire optical system. If indeed the path lengths of the two telescopes to the star are precisely matched, then the combined photometric signal will show a fringe shape with constructive and destructive interference features like the one shown in Figure \ref{fig:fringe}.

How does one get from this observational signal to actually measuring an angular diameter? The short answer to that question is ``by measuring the visibility", which is accomplished by measuring amplitude and phase of the constructive and destructive interference pattern (the fringe; Fig. \ref{fig:fringe}). Every visibility measurement has to be calibrated by removing instrumental and atmospheric systematics. A common practice for interferometric observations is thus the interleaving of observations of target stars with the ones of calibrator stars that are near the target and of similar brightness. Ideally, these calibrators should have known sizes. We therefore use calibrators that are unresolved or at least of smaller angular size than the target, using size estimates from the Jean-Marie Mariotti Center JMDC Catalog\footnote{\url{http://www.jmmc.fr/searchcal} \citep{bon06,bon11,laf10b,laf10a}.}. 

\begin{figure}
\includegraphics[width=\linewidth]{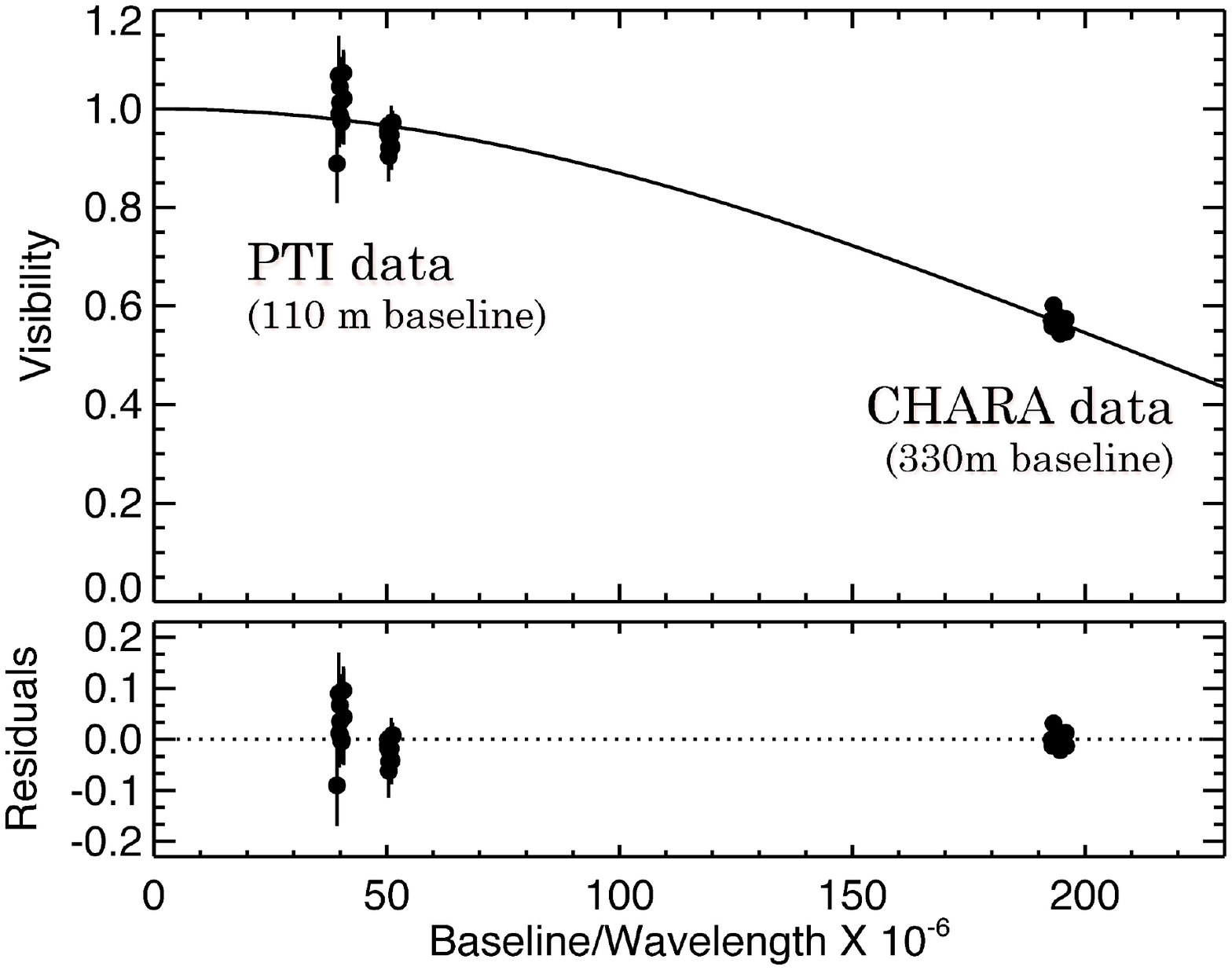}
\caption{Multiple visibility measurements of the star 55~Cancri (\S \ref{chapter:systems:sec:55cnc}) plus the Bessel function fit that corresponds to the Fourier Transform of a stellar disk profile. The interferometric observations were taken at different baselines and interferometers, illustrating how the longer baseline achieves higher spatial resolution and provides more stringent constraints on the fit. Figure adapted from \citet{von11b}. See \S\ref{chapter:parameters:sec:radius} for more details.\label{fig:55Cnc_vis}}
\end{figure}

The measured visibility is dependent upon the (unknown) angular diameter of the star, the (known) projected length of the baseline, and the (known) operational wavelength. The shape of the function relating measured visibility versus baseline (such as in Fig. \ref{fig:55Cnc_vis}) depends upon the topology of the observed object: the Fourier Transform of the object’s brightness distribution perpendicular to the baseline orientation on the sky in the observed wavelength band. This is commonly referred to as the van Cittert-Zernike Theorem \citep{van34,zer38} -- essentially the foundation of astronomical interferometry. 

When extrapolating the situation of doing optical or infrared interferometry with a single baseline to using an infinite number of baselines, one approaches the equivalent of doing astronomical imaging work using a telescope of the size of the longest baseline. Conversely, any two points on the primary mirror of a given telescope can be thought of as two small telescopes in an interferometric array. The image ``seen" by any telescope is equivalent to using an infinite number of 2-telescope arrays and produces, as expected, the topology (the image) of the object under investigation. 

More formally, visibility is defined as the normalized amplitude of the correlation of the light received by two telescopes. It is a unitless number ranging from 0 to 1, where 0 implies no correlation, and 1 implies perfect correlation. An unresolved source would have perfect correlation of 1.0 independent of baseline. A resolved object will show a decrease in visibility toward a value of 0 with increasing baseline length or decreasing wavelength (see Figure \ref{fig:55Cnc_vis}).  The resolvable angular size is proportional to the wavelength of observation and inversely proportional to baseline length. This has the immediate consequence that objects with smaller angular diameters need to be studied at shorter wavelengths and/or larger baselines. 

The brightness distribution of a stellar profile on the sky as seen by an interferometer is a uniform disk, i.e., it does not include the effects of limb darkening due to the decrease in temperature and optical depth toward the outermost layers in the stellar atmosphere (see Figure \ref{fig:mercury}, in which the effects of limb darkening can easily be seen). The visibility dependence upon baseline (Fig. \ref{fig:55Cnc_vis}) may be expressed by $n^{th}$-order Bessel functions, the Fourier Transform of a uniform disk profile, which, as mentioned above, are dependent on the angular diameter of the star, the projected distance between the two telescopes and the wavelength of observation \citep{han74}. 

To convert from the stellar uniform disk diameter, $\theta_{UD}$, to the angular diameter of the Rosseland, or mean, radiating surface of the star, we apply limb-darkening corrections from \citet{cla00,cla11} after we iterate based on the effective temperature value obtained from initial spectral energy distribution fitting (see \S \ref{chapter:parameters:sec:temp})\footnote{Limb-darkening coefficients are dependent on assumed stellar effective temperature, surface gravity, and weakly on metallicity. When we vary the input $T_{\rm eff}$ by 200 K and $\log g$ by 0.5 dex, the resulting variations are below 0.1\% in $\theta_{\rm LD}$ and below 0.05\% in $T_{\rm eff}$. Varying the assumed metallicity across the range of our target sample does not influence our final values of $\theta_{\rm LD}$ and $T_{\rm eff}$ at all.}. 

The limb-darkening corrected angular diameters, $\theta_{LD}$, coupled with trigonometric parallax values from \citet{van07}, determine the linear stellar diameters. Uncertainties in the physical stellar radii of main-sequence stars are typically dominated by the uncertainties in the angular diameters, not the distance. For giant stars that are at larger distances, however, distance uncertainties can play a more significant role in the calculation of the total error budget. 

It is worth pointing out the awesome resolving power of interferometers such as CHARA. We typically achieve 1--3\% precision on stellar angular diameters that are a fraction of a milliarcsecond in size. To visualize this: a soccer ball on the moon, the width of a human hair at 10 miles distance, or the size of a coin in Australia as seen from the USA, approximately correspond to an angular size of one milliarcsecond. 

We note that interferometry can be prone to systematic errors that arise in the process of correcting the data for atmospheric effects. It is thus very difficult to properly assess interferometric uncertainties \citep{boy13,hub16}. Calibrator stars should ideally be point sources (i.e., have a diameter of zero size or at least one that is much smaller than the resolution of the interferometer) and should be observed on time scales equal to or smaller than the typical time scale of a change in atmospheric conditions. These assumptions cannot always be met. Calibrators need to be close in angular distance from the target and cannot be much brighter or fainter, meaning that one could be forced to pick calibrators with non-zero angular diameters or even diameters  comparable with the one of the target itself. To properly extract the target data, however, we need to assume  calibrator diameters, which may not be accurately known. This situation is, somewhat ironically, exacerbated by the continuously improving resolving and light gathering powers of interferometers -- one simply runs out of unresolved calibrators.  

Furthermore, there are two time scales on which conditions in the atmosphere change. The shorter one of these is the coherence time, which is less than a second in duration \citep[e.g., ][]{kel07}. High-frequency fringe tracking or fringe scanning is used to freeze the atmospherice conditions during exposures, which can be challenging from an engineering and software stand point. The longer time scale is the one over which the atmosphere changes in a more global way, i.e., over which measured calibrator visibilities will vary. This time scale is dependent on the stability of the atmosphere and can be minutes to hours long. To correct for this effect, we employ the aforementioned alternating observations of target and calibrators, which takes minutes to a few tens of minutes. 

Consequently, $\chi^{2}_{reduced}$ values of the visibility fits, such as shown in Figure \ref{fig:55Cnc_vis},   may be calculated to be smaller than unity due to the aforementioned difficulties of accurately defining uncertainties in visibility measurements. While there are mathematical methods of tracking errors through the calibration of visibility via standard statistical methods \citep[e.g.,][]{van05}, the atmospheric effects described earlier introduce effects that cannot be calculated via those or similar methods. Therefore, we set a true $\chi^{2}_{reduced}$ = 1 and calculate uncertainties for $\theta_{\rm UD}$ and $\theta_{\rm LD}$ based on a rescaling of the associated uncertainties in the visibility data points. That is, the estimates of our uncertainties in $\theta_{\rm UD}$ and $\theta_{\rm LD}$ are based on a $\chi^{2}_{reduced}$ fit, not on strictly analytical calculations, though see the recent publication by \citet{nun17} for a new analytical/empirical approach of calculating uncertainty values.

For much greater detail on the technique of interferometry in astronomy, we refer the reader to \citet{law00, mon03, ten13a, ten13b, lab14, van15, bus15} and references therein.

 
\section{Stellar Effective Temperature and Luminosity}
\label{chapter:parameters:sec:temp}

From a combination of the Stefan-Boltzmann Law and the formula relating angular to physical stellar diameter based on distance, it is trivial to derive

\begin{equation} \label{eq:temperature}
T_{\rm eff} ({\rm K}) = 2341 (F_{\rm BOL}/\theta_{\rm LD}^2)^{\frac{1}{4}}. 
\end{equation}

\noindent
In the above equation, $T_{\rm eff}$ is the stellar effective temperature and is defined as the surface temperature of a black body that emits the same amount of energy per unit time as the star. It typically serves as a measure of stellar surface temperature when the stellar emissivity as a function of wavelength is unknown. $F_{\rm BOL}$ is the bolometric flux as measured by a hypothetical detector across all wavelengths; it is in units of $10^{-8}$~erg cm$^{-2}$ s$^{-1}$. $\theta_{\rm LD}$ is the limb-darkening corrected stellar angular diameter (see \S \ref{chapter:parameters:sec:radius}) in units of milliarcseconds (mas).

We can see from Equation \ref{eq:temperature} that, to obtain direct measurements of stellar effective temperature and luminosity, we need stellar angular diameter (e.g., from interferometry; \S \ref{chapter:parameters:sec:radius}) and bolometric flux. $F_{\rm BOL}$ is typically determined by spectral energy distribution (SED) fitting: determining the integral of received spectral flux density over all wavelengths. 

\begin{figure*}
\includegraphics[width=\linewidth]{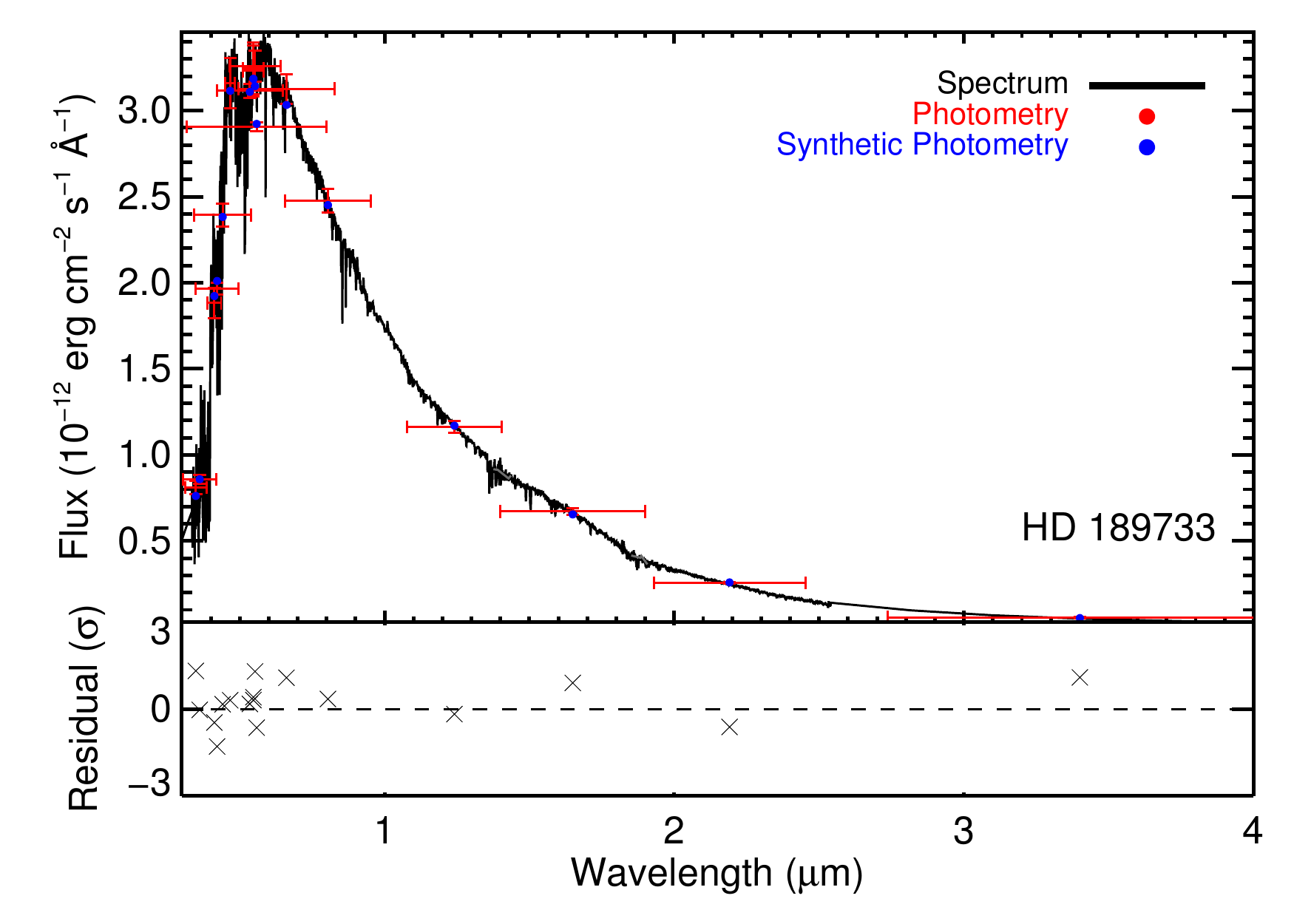}
\caption{SED fit for HD~189733 to illustrate our SED fitting routine when using spectrophotometry data. The black spectrum shows spectrophotometry data of HD~189733 that are flux calibrated modulo a zero point offset. The red points indicate literature photometry data to determine this zero point offset, along with the width of the filters shown as horizontal bars. The blue points represent the flux value of the spectrophotometry data integrated over the filter transmission profile. The lower panel displays the residuals (literature data minus spectral template) in units of standard deviation of the literature photometry data points. Adapted from \citet{boy15}. For more details, see \S\ref{chapter:parameters:sec:temp}.
\label{fig:hd189733_sed}}
\end{figure*}

In order to stay as model-independent as possible in our SED fitting approach, we ideally use spectrophotometry data of the star under investigation and scale it to flux-calibrated literature photometry. The principal advantage of such spectrophotometric data is that they show all spectral features of the star under investigation across the full range of wavelength coverage, which is particularly relevant for SED fitting accuracy of low-mass stars. An example of this is shown in Figure \ref{fig:hd189733_sed}. The black spectrum shows spectrophotometry data of HD~189733 (\S \ref{chapter:systems:sec:hd189733}) that are flux calibrated modulo a zero point offset. The red points indicate literature photometry data to determine this zero point offset, along with the width of the filters shown as horizontal bars. The blue points represent the flux value of the spectrophotometry data integrated over the filter transmission profile.

\begin{figure*}
\includegraphics[width=\linewidth]{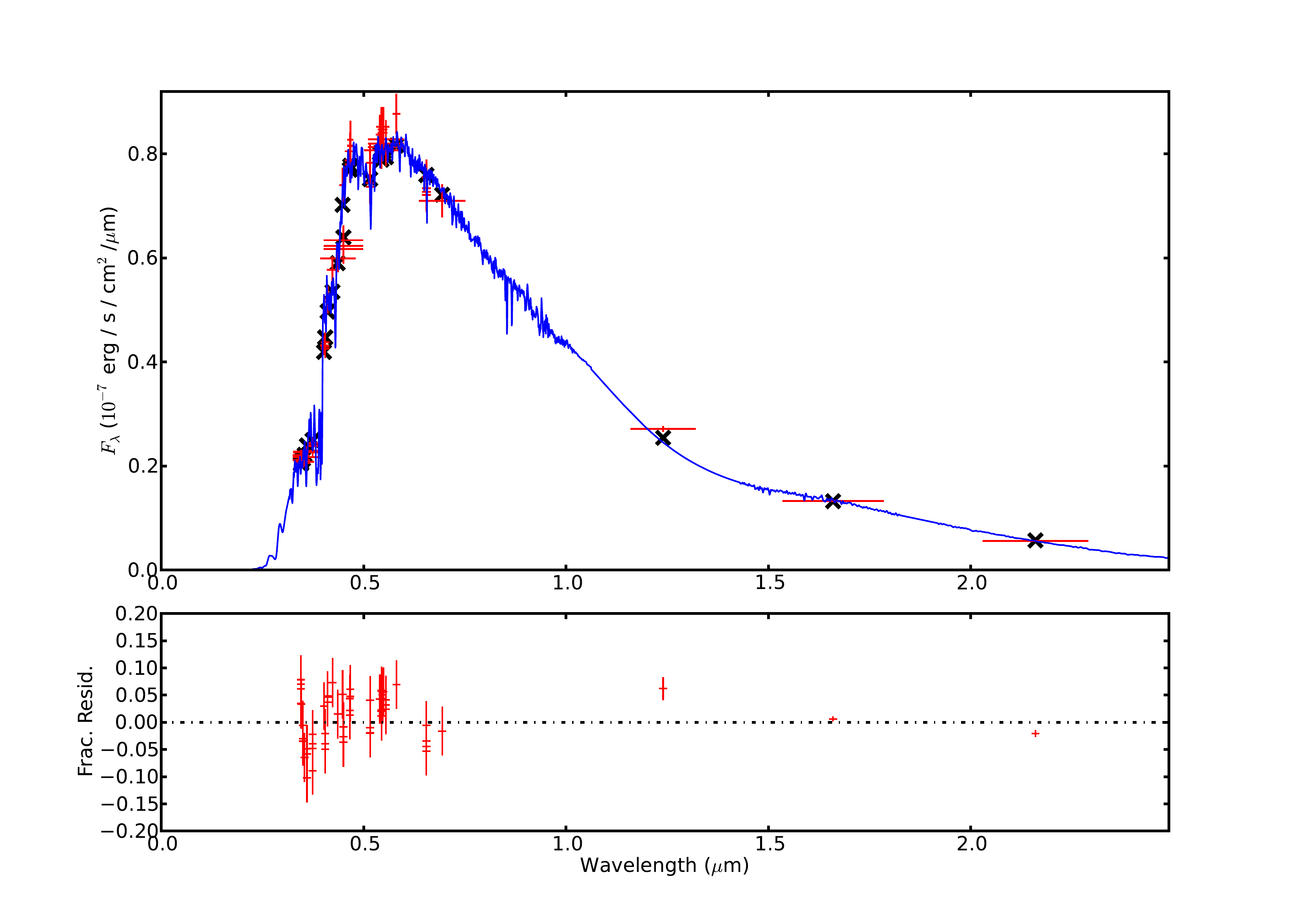}
\caption{SED fit for GJ~614 to illustrate our SED fitting routine when using spectral templates. The blue spectrum shows the \citet{pic98} template for a G9IV star. The red crosses indicate literature photometry data and filter width as horizontal bars, and the black crosses show the flux value of the spectral template integrated over the filter transmission profile. The lower panel displays the residuals (literature data minus spectral template) in units of standard deviation of the literature photometry data points. Adapted from \citet{von14}. For more details, see \S\ref{chapter:parameters:sec:temp}.
\label{fig:gj614_sed}}
\end{figure*}

If no spectrophotometry data are available, we fit spectral templates from \citet{pic98} or optical spectra of the target star \citep[e.g., ][]{man13} to flux-calibrated literature photometry or spectrophotometry. One such example is illustrated in Figure \ref{fig:gj614_sed} for the star GJ~614. The blue spectrum shows the \citet{pic98} template for a G9 subgiant. The red crosses indicate literature photometry data values as well as filter bandwidth as horizontal bars. The black crosses indicate the flux value of the spectral template integrated over the filter transmission profile.

In both cases, either the spectral template or the spectrophotometric data are scaled to literature data to minimize $\chi^2$ and subsequently integrated over wavelength to obtain the bolometric flux. 

Our SED fitting strategy furthermore includes the following aspects: 
\begin{itemize}

\item Extinction is a variable that can be set to float or to zero for nearby targets. Since most exoplanet hosts and main-sequence stars for which we can achieve the necessary angular resolution are very close, extinction is in fact set to zero for most of our targets \citep{boy13,boy14,von14}. 
\item Some literature photometry data have no quoted uncertainty associated with it, particularly older data sets. We assign a 5\% random uncertainty to any such data point for the calculation of the fit's $\chi_{reduced}^2$.
\item The use of literature spectrophotometry in addition to broad-band photometry may increase the fit's $\chi_{reduced}^2$. Nonetheless, we include spectrophotometry data whenever available in order to reduce any potential systematics in the choice of spectral template.

\end{itemize}

Below, we point out some recent modifications and improvements to our SED fitting approach. 
\begin{itemize}
\item Calculated uncertainties in the bolometric flux values are statistical only. A posteriori, it is impossible to account for and correct possible systematics in the literature photometry such as saturation or correlated errors, filter errors due to problems with transmission curves \citep{man15a}, or other non-random error sources. In order to provide a quantitative estimate for these unknown errors, the detailed study by \citet{boh14} recommends inflating random errors in $F_{\rm BOL}$ by 2\% in quadrature. We incorporate this additional random error into our measurements starting with \citet{boy15} and all subsequent publications.
\item Initially during our SED fitting, the filters of the literature photometry data were assumed to have a top-hat shape \citep{van09,von11c,von11a,von12,von14,boy12a,boy12b,boy13}. That is, during the calculation of $\chi^2$, only the central filter wavelengths were correlated with the SED template's flux value averaged over the filter transmission range in wavelength. The study by \citet{man15a} motivates the modification of actual filter curves to correct a posteriori for unknown errors in the literature photometry. Their calibration is based on integration of space-based, flux-calibrated spectrophotometry across broad-band filters. With small, parametrized adjustments to the shape of broad-band filter transmission curves, \citet{man15a} accomplish consistent results throughout the literature. As with the additional random error described above, we incorporate the \citet{man15a} technique into our SED fitting routine starting with the \citet{boy15} publication and all subsequent papers.  
\end{itemize}

During the SED fitting procedure, the only systematics that {\it can} be controlled are (1) the choice of spectral template for the SED fit, (2) the choice of which literature photometry data points to include in the fit, and, to a lesser extent, (3) whether to let the interstellar reddening float during the fit or whether to set it to zero. 


\section{Why Interferometry?}
\label{chapter:parameters:sec:why}

Equation \ref{eq:temperature} shows the inter-dependence between stellar angular diameter, effective temperature, and bolometric flux. The approach outlined above involves obtaining the angular diameter interferometrically, measuring bolometric flux based on literature photometry, and calculating effective temperature. Conducting interferometric observations, however, is not trivial, there are very few places to do it, and limits in stellar brightness and angular size may reduce target availability. A fair question to be asked is why one would not choose to invert the problem: obtain effective temperature and bolometric flux some other way, and calculate stellar angular diameter, which is then trivially converted to physical radius with the knowledge of distance \citep[e.g., ][]{man13,man15b}.

The short answer is that the precision, and most likely accuracy as well, is greater when using interferometry, provided the star under investigation is observable interferometrically; otherwise, the approach outlined above is certainly the best way to go. To look at this statement in more detail, let's consider two different scenarios: (1) bolometric flux of the target star is measurable and effective temperature can be obtained from semi-empirical studies such as the infrared flux method \citep{cas10}, or (2) stellar bolometric flux is not known, and the use of stellar models, for instance based on input spectra, is required to fully characterize the target.

From Equation \ref{eq:temperature}, it is readily calculable that the relative uncertainty in angular diameter is twice that of the relative uncertainty in the effective temperature for non-systematic errors. Thus, a 3-5\% error in the determination of effective temperature would result in an uncertainty in calculated angular diameter of 6-10\%. In comparison, interferometric errors in the angular diameters for the majority of our published work are on the order of 1-3\%. 

Stellar models on other hand, though hugely important in all of stellar astrophysics, thus far continue to produce discrepant results for stellar diameters (too small) and effective temperatures (too hot) when compared to directly determined values (\S \ref{chapter:systems}). These offsets vary from star to star but are generally larger for stars with spectral types much later or much earlier than the Sun. The canonical values for KM dwarfs are around 5\% and 3\% for stellar diameter and effective temperature, respectively \citep[e.g.,][]{man15b}, but as we point out in Sections \ref{chapter:systems:sec:gj876} and \ref{chapter:systems:sec:gj436} for instance, they can be much higher. Moreover, the question of accuracy remains. Obtaining effective temperatures from spectroscopic studies involves making assumptions, like masking out bad regions in the stellar spectra \citep{man13}, knowing or assuming the metallicity and its effect on opacities \citep{new14,man15b}, breaking the degeneracy between spectral signatures of $\log g$, [Fe/H], and $T_{\rm eff}$ \citep{buz01}, etc. 

These two scenarios provide motivation of interferometry as a method to study stellar parameters in spite of its limitations. In fact, spectroscopic and/or other semi-empirical $T_{\rm eff}$ determinations and stellar models tend to rely on interferometric data as calibrations \citep[see for example][]{spa17}. Finally, interferometry will be clearly superior in its reliability when involving more exotic stars, such as young, very cool, or very hot stars, very metal-rich or metal-poor, etc., for which data do not exist at an adequate level to provide anything but weak constraints on the corresponding models.


\section{System Habitable Zone}
\label{chapter:parameters:sec:hz}

\begin{figure}
\includegraphics[width=\linewidth]{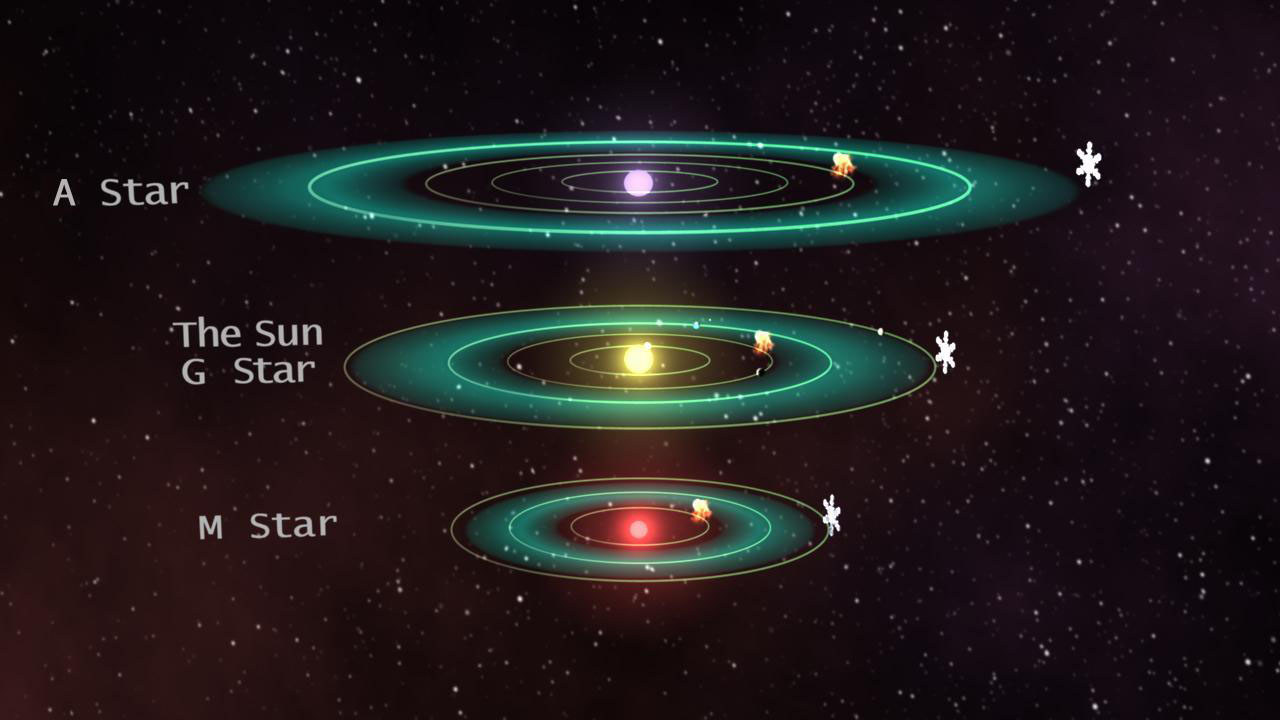}
\caption{The basic concept of a stellar HZ. Shown are main-sequence stars of three different spectral types. The earlier the stellar spectral the further out and wider the system HZ. Image credit: NASA / NASA Astrobiology Institute. \label{fig:hz}}
\end{figure}

The habitable zone of a planetary system is described as the range of distances at which water on a planet with a surface and an atmosphere containing a modest amount of greenhouse gases would be in liquid phase. The concept was first characterized in \citet{kas93}, and there have since been a number of extensions and refinements of that formalism. The basics of the concept are shown in Fig. \ref{fig:hz}. The hotter the star, the farther out the system HZ is located. 

It is worth noting that there is a big difference between the concepts of inhabited planets, habitable planets, and planets in the habitable zones of stars. Of the first two categories we know, at least at the time of this writing, of only one -- Earth. But there are many known exoplanets in habitable zones. To constrain any  probability of habitability of a planet in an HZ, at least with respect to life somewhat akin to the one we know, one would have to consider the following aspects: (1) does the planet have a surface on which there could be liquid water? (2) does the planet have liquid water on that surface? (3) is the planetary atmosphere such that the temperature at the surface is not too hot or too cold for there to be water in liquid phase, i.e., are there the correct amounts of greenhouse gases in the planetary atmosphere? (4) does the planet have a magnetic field that would protect any life on its surface from incident solar wind? (5) is the parent star sufficiently quiescent so as to not obliterate any evolving life with massive flares? 

Of course, the above aspects are qualitative, rely on many assumptions, and the list is not complete. For instance, the planet would also have to be in a dynamically stable orbit, i.e, not be ejected out of or moved within the stellar system by perturbing objects, and in an orbit that is not too elliptical so as to cause massive temperature variations on the surface. Additionally, the existence of a moon would stabilize the planet's axis of rotation. The principal point is that HZ locations and boundaries can be scientifically calculated, but when any discussion ventures from habitable zone to habitability, many other aspects come into play that are between very difficult and impossible to determine. 

\citet{kop13a,kop13b,kop14} provide an online calculator\footnote{http://www3.geosc.psu.edu/$\sim$ruk15/planets/ or http://depts.washington.edu/naivpl/content/hz-calculator} for HZs, thereby defining the HZ boundaries based on a runaway greenhouse effect (too hot) or a runaway snowball effect (too cold) as a function of stellar luminosity and effective temperature, plus water absorption by the planetary atmosphere. Whichever assumption is made of how long Venus and Mars were able to retain liquid water on their respective surfaces defines the choice of HZ (conservative or optimistic). These conditions are described in more detail in \citet{kop13b}, section 3 of \citet{kan13}, and section 2 of \citet{kan16}\footnote{Note that the concept of a HZ around multiple stars, referred to as the circumbinary HZ, is explained in, e.g., \citet{kan13b}. For the purpose of this publication, we will only use the HZ around single stars.}. 

The system HZ is dependent on both the stellar luminosity and effective temperature, both obtained empirically by the interferometric studies motivated here. The dependence of habitable zone distances and width upon stellar parameters plus associated uncertainties is quantitatively addressed in \citet{kan14} and \citet{cha16}, including Gaussian error propagation in section 5.5 of \citet{cha16}. It is of vital importance to have one's stellar astrophysical parameters correct when making, e.g., statistical statements of frequency of HZ planets or fraction of stars with Earth-sized planets in the HZ ($\eta_{\oplus}$). Note that the latter is of particular significance for late-type dwarfs since (1) the short-period sensitivity of transit and RV surveys makes the determination of $\eta_{\oplus}$ for this population easier than for hotter stars, but (2) the divergence between calculated and measured stellar parameters is highest for late-type stars. The need to harmonize stellar model predictions with empirical data products is thus clearly illustrated when describing HZ regions for current and upcoming targets, such as the ones relevant for the upcoming Transiting Exoplanet Survey Satellite \citep[TESS, ][]{ric14}.


\section{Transiting Planets}
\label{chapter:parameters:sec:transiting}

\begin{figure}
\includegraphics[width=\linewidth]{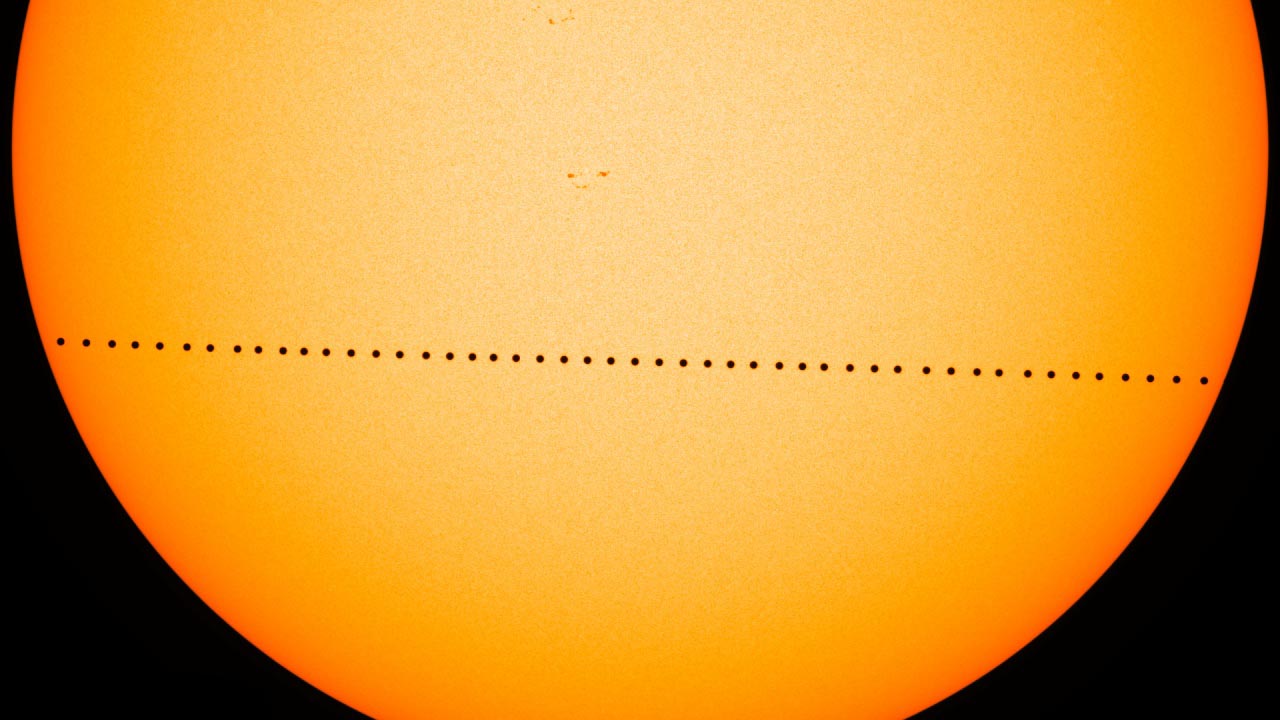}
\caption{The concept of a transiting planet illustrated by a composite, visible light image of Mercury's transit across the sun in 2016. Also clearly visible in this image is the effect of limb darkening: the star's surface brighness is lower toward the limb. For discussion on limb darkening, see Section \ref{chapter:parameters:sec:radius}, and for transiting planets, see Section \ref{chapter:parameters:sec:transiting}. Image credit: NASA / Solar Dynamics Observatoty. \label{fig:mercury}}
\end{figure}

Transiting planets are particularly valuable for the study of exoplanets because the inclination of the plane of the system with respect to the line of sight is known. If stellar RV variations due to the planet can be measured, the determination of actual planet mass, as opposed to minimum planet mass, becomes possible. Furthermore, the planetary radius can be determined from the measured flux decrement and knowledge of, or assumption about, stellar radius \citep[see][for details and equations]{win10}. Figure \ref{fig:mercury} shows a composite image of Mercury's transit across the face of the sun in 2016 as captured by NASA's Solar Dynamics Observatory. At this point, there has not yet been a successful attempt to interferometrically resolve a transit signature across a star other than the sun due to signal-to-noise issues \citep[for details, see][]{van08a}, but the flux decrement is readily observed photometrically.

Without any direct measurements, this stellar radius has to be either derived from stellar models, or from assumptions about the stellar mass coupled with a mass-radius relation or coupled with the analysis of the transiting planet's light curve \citep{tor07}.

If the stellar radius can be measured, however, then the planetary radius and bulk density are directly calculable from the flux decrement (see \S\S \ref{chapter:systems:sec:55cnc}, \ref{chapter:systems:sec:hd209458}, and \ref{chapter:systems:sec:hd189733}; also, see \citealt{win10}). Moreover, a comprehensive analysis involving literature light curves and RV curves will furthermore allow for the calculation of stellar parameters such as mass, density, and surface gravity (see \S \ref{chapter:systems:sec:gj436}). 


%
%
%
\chapter{Results}
\label{chapter:systems} 

\newcommand{\one}{HD\,189733}	
\newcommand{\two}{HD\,209458}	
\newcommand{\thetaone}{$0.3848 \pm 0.0055$}	
\newcommand{\thetatwo}{$0.2254 \pm 0.0072$}	
\newcommand{\thetaoneud}{$0.3600 \pm 0.0046$}	
\newcommand{\thetatwoud}{$0.2147 \pm 0.0066$}	

\newcommand{\fboloneorig}{$2.785 \pm 0.015$}		
\newcommand{\fboltwoorig}{$2.331 \pm 0.020$}		
\newcommand{\fbolone}{$2.785 \pm 0.058$}		
\newcommand{\fboltwo}{$2.331 \pm 0.051$}		
\newcommand{\luminone}{$0.328\pm0.011$}	
\newcommand{\lumintwo}{$1.794\pm0.148$}		

\newcommand{\rhoone}{$1.62 \pm 0.11$}	
\newcommand{\rhotwo}{$0.58 \pm 0.14$}	
\newcommand{\rhoonep}{$0.605 \pm 0.029$}	
\newcommand{\rhotwop}{$0.196 \pm 0.033$}	
\newcommand{\rhoonecgsp}{($0.802 \pm 0.038$)~g~cm$^{-3}$}	
\newcommand{\rhotwocgsp}{($0.260 \pm 0.043$)~g~cm$^{-3}$}	

\newcommand{\radonep}{$1.216 \pm 0.024$}	
\newcommand{\radtwop}{$1.451 \pm 0.074$}	
\newcommand{\radone}{$0.805 \pm 0.016$}	
\newcommand{\radtwo}{$1.200 \pm 0.061$}	

\newcommand{\teffone}{$4875 \pm 43$}	
\newcommand{\tefftwo}{$6098 \pm 101$}	

\newcommand{\loggone}{$4.56 \pm 0.03$}	
\newcommand{\loggtwo}{$4.28 \pm 0.10$}	
\newcommand{\loggonep}{$3.29 \pm 0.02$}	
\newcommand{\loggtwop}{$2.88 \pm 0.07$}	



In this Chapter, we provide a table with the currently available high-precision, interferometrically measured stellar angular diameters  (\S~\ref{chapter:systems:sec:status}), elaborate on a number of individual exoplanetary systems that were characterized by means of interferometric studies (\S~\ref{chapter:systems:sec:individual}), and briefly comment on our overall results (\S~\ref{chapter:systems:sec:summary}). 


\section{Current Status}
\label{chapter:systems:sec:status}

Table \ref{tab:systems:dwarfs} provides a summary of all stellar systems with interferometrically determined angular diameters with precision better than 5\%, smaller than 100 solar radii, out to a distance of 150pc, as of 1 Nov 2016. We exclude fast rotators and stars with large pulsations as one cannot assume a respective single value for the angular diameter in these cases. Except where indicated, we use the respective publication's values for stellar angular diameter and bolometric flux, plus trigonometric parallax from either the publication itself or \citet{van07}, to uniformly calculate physical stellar radii, effective temperatures, and luminosities. 
If no uncertainty for a physical parameter was given, calculated values of derived quantities do not have any associated uncertainty, either. Spectral types are for reference only and are taken from \citet{and12}. 

\begin{figure}
\includegraphics[width=\linewidth]{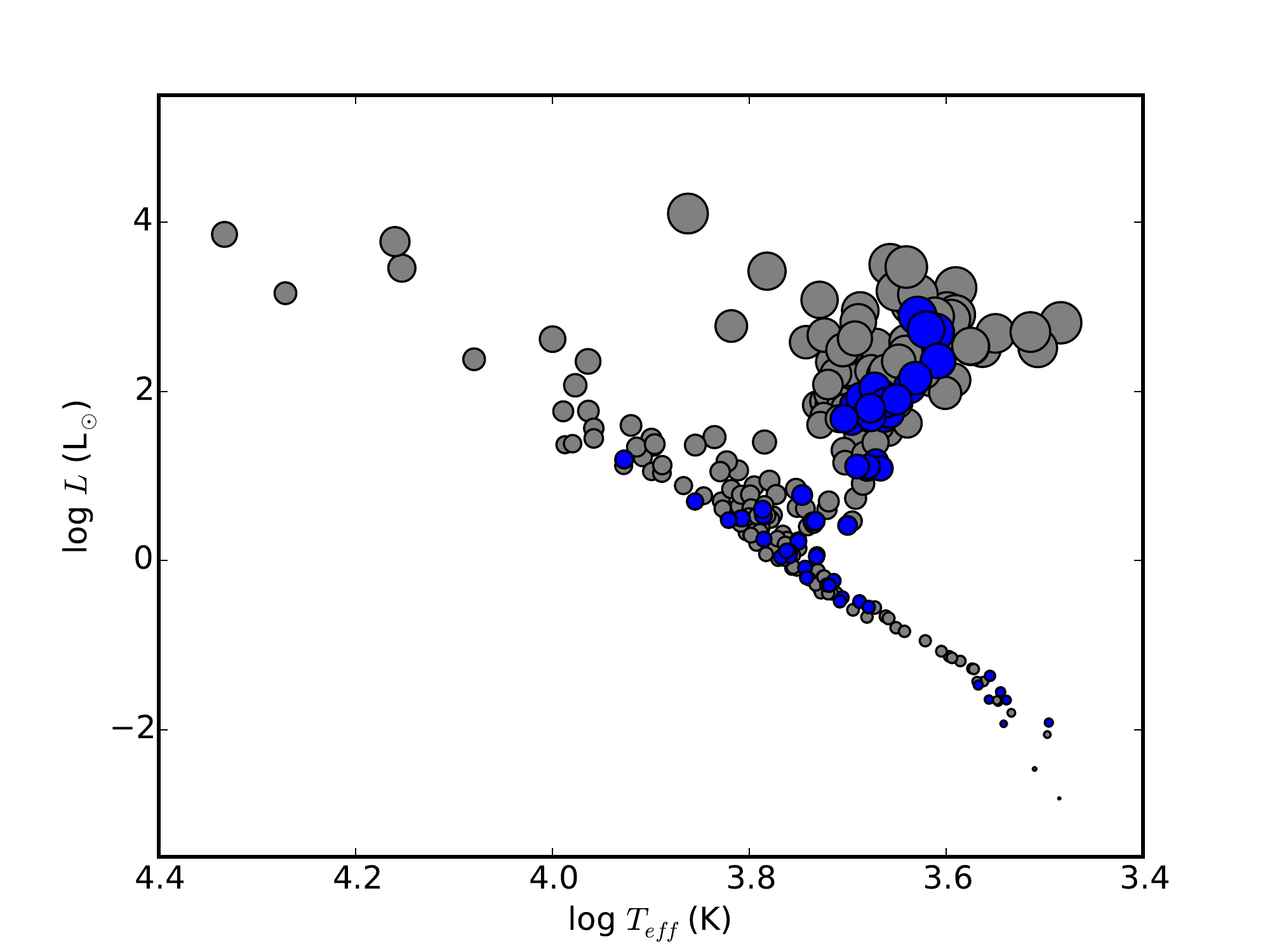}
\caption{Empirical H-R Diagram for all stars with interferometrically determined stellar radii with random uncertainties smaller than 5\%, stellar radii $<$ 100 \rsun, and distances up to 150 pc (status 1 Nov 2016). Stellar effective temperature, luminosity, and radius data for this figure are taken from Table \ref{tab:systems:dwarfs}. The diameter of each data point is representative of the logarithm of the corresponding stellar radius. Error bars in effective temperature and luminosity are smaller than the size of the data points. Stars shown as blue circles are known to host exoplanets, stars shown as grey circles are not. See also Fig. \ref{fig:HRD_allstars}, which is based on the same data. For more information, see Section \ref{chapter:systems:sec:status}.
\label{fig:HRD_EHS_blue}}
\end{figure}

The last column contains a list of numeric reference codes for the angular diameter value(s), an letter reference code for value(s) of bolometric flux, and a ``YES" in case the system is known to host at least one extrasolar planet. The reference codes are given below. Wherever more than one reference is given, the values for angular diameter and/or bolometric flux and associated uncertainties are calculated via weighted averages from which stellar radii, temperatures, and/or luminosities are subsequently derived. Figure \ref{fig:HRD_EHS_blue} graphically depicts the contents of Table \ref{tab:systems:dwarfs} and identifies exoplanet hosts in blue versus stars not currently known to host planets in orbits in grey. The range in radii extends from about 0.15 \rsun to 100 \rsun. 


\begin{center}
\begin{small}
\begin{landscape}

\clearpage
\end{landscape}

\end{small}
\end{center}





\noindent{\bf References for stellar angular diameter data: }

\noindent 
1: \citet{boy13}; 
2: \citet{boy12b}; 
3: \citet{ber06}; 
4: \citet{von14}; 
5: \citet{mae13}; 
6: \citet{bai08}; 
7: \citet{lig12}; 
8: \citet{boy12a}; 
9: \citet{boy08}; 
10: \citet{dif04}; 
11: \citet{van09}; 
12: \citet{dif07}; 
13: \citet{the05}; 
14: \citet{dem09}; 
15: \citet{hen13}; 
16: \citet{dav11}; 
17: \citet{han74}; 
18: \citet{dav86}; 
19: \citet{ker03a}; 
20: \citet{moz03}; 
21: \citet{big11}; 
22: \citet{chi12}; 
23: \citet{nor01}; 
24: \citet{ker04a}; 
25: \citet{von11a}; 
26: \citet{lan01}; 
27: \citet{von12}; 
28: \citet{cre12}; 
29: \citet{big06}; 
30: \citet{ker03b}; 
31: \citet{bai13b}; 
32: \citet{baz11}; 
33: \citet{hub12}; 
34: \citet{per13}; 
35: \citet{whi13}; 
36: \citet{nor99}; 
37: \citet{boy15}; 
38: \citet{crep12}; 
39: \citet{ker08}; 
40: \citet{seg03}; 
41: \citet{bai12}; 
42: \citet{von11b};
43: \citet{tan15};
44: \citet{boy09};
45: \citet{bai10b};
46: \citet{bai09};
47: \citet{mer10};
48: \citet{bai11a};
49: \citet{der11};
50: \citet{bai11b};
51: \citet{sim11};
52: \citet{bai13a};
53: \citet{joh14};
54: \citet{cre15};
55: \citet{des08};
56: \citet{vcb07};
57: \citet{cia01};
58: \citet{van09b};
59: \citet{van99};
60: \citet{cus12};
61: \citet{1999AJ....118.3032N};
62: \citet{2003AJ....126.2502M};
63: \citet{2001AJ....122.2707N};
64: \citet{1991AJ....101.2207M};
65: \citet{1989ApJ...340.1103H};
66: \citet{1993A&A...270..315D};
67: \citet{bro74};
68: \citet{2004A&A...413..711W};
69: \citet{1992AJ....104.2224R};
70: \citet{2000ApJ...543..972N};
71: \citet{1987AJ.....94..751W};
72: \citet{2005A&A...433..305R};
73: \citet{2004A&A...421..703W};
74: \citet{1996AJ....111.1705D};
75: \citet{1992AJ....104.1982D};
76: \citet{1996ApJ...463..336B};
77: \citet{1988ApJ...327..905S};
78: \citet{dav86};
79: \citet{2001A&A...377..981W};
80: \citet{1996A&A...312..160Q};
81: \citet{1986A&A...166..204D};
82: \citet{1995AJ....109..378D};
83: \citet{1997AJ....114.2150V};
84: \citet{2003SPIE.4838..369Y};
85: \citet{1991AJ....102.2091B};
86: \citet{2001ApJ...559.1147C};
87: \citet{2004ApJ...612..463O};
88: \citet{2003IAUS..219E.127K};
89: \citet{lig16};
90: \citet{bai16};
91: \citet{ker16}. 
\\
\noindent{\bf References for the $F_{BOL}$ measurements: }

\noindent 
a: \citet{boy13}; 
b: \citet{man13}; 
c: \citet{von14}; 
d: \citet{mae13}; 
e: \citet{boy12b}; 
f: \citet{hen13}; 
g: \citet{dav11}; 
h: \citet{bai13b}; 
i: \citet{egg08}; 
j: \citet{hub12}; 
k: \citet{boy15};
l: \citet{tan15};
m: \citet{and11};
n: \citet{boy09};
o: \citet{bai10b};
p: \citet{bai09};
q: \citet{bai11a};
r: \citet{bai11b};
s: \citet{cre12};
t: \citet{bai13a};
u: \citet{joh14};
v: \citet{cre15};
w: \citet{cod76};
x: \citet{alo94};
y: \citet{van09b};
z: \citet{van99};
aa: \citet{man15b};
ab: \citet{lig16};
ac: \citet{bai16}.


\section{Selected Individual Systems}
\label{chapter:systems:sec:individual}

The following Sections describe some individual exoplanetary systems that were studied with interferometry. They are somewhat arbitrarily chosen with the aim to illustrate recent progress in the field, show particularly interesting systems from a scientific or historic point of view, and emphasize insights gained by obtaining direct stellar radii and effective temperatures. 


\subsection{51~Pegasi}
\label{chapter:systems:sec:51peg}

The discovery of an exoplanet around the star 51~Peg (HD~217014) is widely considered to have ushered in the era of extrasolar planet discoveries \citep{may95}, thereby surprising the community by its system architecture that is so different from the solar system. While it heralded the birth of observational exoplanetary science, the system itself turned out to be one of many similar subsequent discoveries: a hot Jupiter in orbit around a nearby \citep[$d=15.6$pc;][]{van07}, sun-like star. The planet does not transit its parent star, but the star is readily accessible via interferometry from CHARA and other facilities. The stellar radius was first measured by \citet{bai08}, confirmed (albeit at lower precision) by \citet{van09}, and updated by \citet{boy13}: R = $(1.137\pm0.015) R_{\odot}$, $T_{\rm eff}$ = (5781$\pm$36) K, L = $(1.301\pm 0.016) L_{\odot}$ (see Table \ref{tab:systems:dwarfs}).


\subsection{GJ~581}
\label{chapter:systems:sec:gj581}

GJ~581 (HIP~74995) is a nearby \citep[$d=6.3$pc;][]{van07} early M dwarf \citep{bes90a,hen94,haw96,cut03} that hosts multiple reported planets, all detected by the radial velocity technique \citep{bfd05,udr07,mbf09,vog10a}. The exact number of planets in orbit around this star is a matter of debate in the astronomical literature, ranging from three to six. 

\begin{figure*}		
	\begin{center}								
    \includegraphics[width=\linewidth]{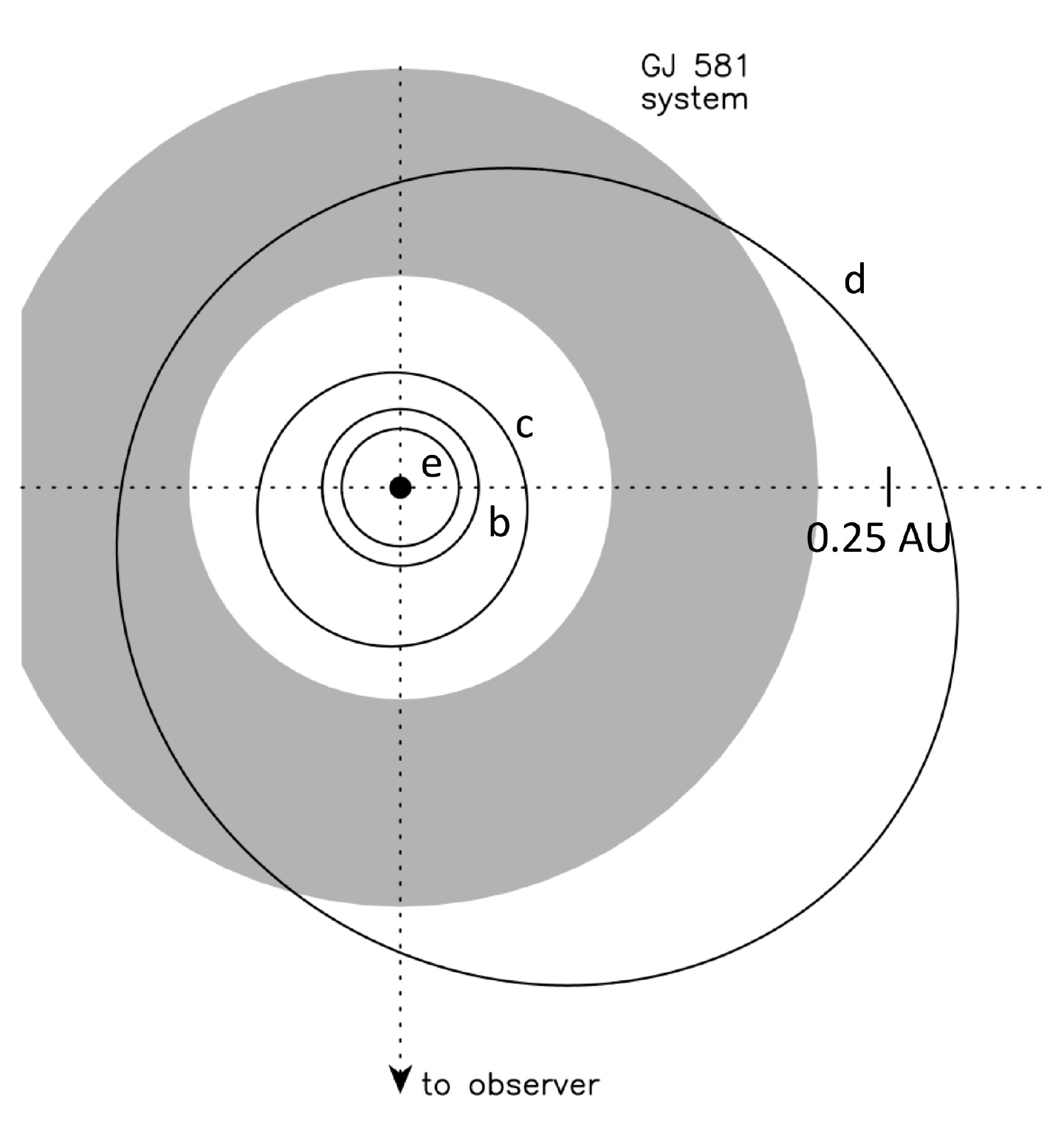} 	
  \caption{A top-down view of the GJ~581 system based on the \citet{mbf09} orbital elements. The habitable zone is indicated by the gray-shaded region with calculated boundaries of 0.11 and 0.21 AU. Planet d spends part of its elliptical orbit in the HZ. Note that other orbital architectures are published in, e.g., \citet{vog10a,ang10}. Planet f ($a$ = 0.758 AU) is not shown for purpose of clarity. For more details, see \S \ref{chapter:systems:sec:gj581}. Figure adapted from \citet{von11a}.}
  \label{fig:hz_orbits}
  \end{center}
\end{figure*}

Based on \cite{von11a} and \citet{man13}, GJ~581's stellar parameters are given in Table \ref{tab:systems:dwarfs} to be R = $(0.299\pm 0.01) R_{\odot}$, $T_{\rm eff}$ = (3480$\pm$55) K, and L = $(0.0118 \pm 0.0002) L_{\odot}$. Following the arguments in \cite{von11a} that are based on the equations in \citet{und03} and \citet{jon10}, GJ~581's HZ can be calculated to have an inner boundary of 0.11~AU and an outer boundary of 0.21~AU -- see also Section \ref{chapter:parameters:sec:hz} on the calculation of HZ boundaries. Since the orbital elements are known from the dynamical studies of the planetary system, even though they may differ depending on the respective publication (e.g., \citealt{mbf09} vs \citealt{vog10a,ang10}), the determination of which planets spend how much of their orbits inside the system HZ is straightforward -- see Fig. \ref{fig:hz_orbits}. 

We can use the stellar parameters to calculate the planetary equilibrium temperatures $T_{eq}$ following the methods of \citet{skl07}:
\begin{equation}\label{eq:equitemp}
  T_{eq}^4 = \frac{S (1 - A)}{f \sigma},
\end{equation}
where $S$ is the stellar energy flux, $A$ is the Bond albedo, and $\sigma$ is the Stefan-Boltzmann constant \citep{skl07}. In this definition, the redistribution factor $f$ is determined by the efficiency of atmospheric heat redistribution efficiency and is set to 2 for a hot dayside and much cooler nightside, and it is set to 4 for even heat redistribution, i.e., similar temperatures between dayside and nightside. Note that although a planet's orbit may be located in the system HZ, its equilibrium temperature may be below the freezing point for water (e.g., Earth's $T_{eq} = 255$~K), since the $T_{eq}$ values do not take into account the greenhouse effect \citep{wor10}. See \citet{von11a} for more details.

Figure \ref{fig:hz_orbits} depicts one of the published architectures, in this case the one from \citet{mbf09}, of the GJ~581 system with its HZ shown as a gray-shaded region\footnote{Note that the concept of optimistic versus conservative HZ boundaries did not exist until 2013 (see \S \ref{chapter:parameters:sec:hz}).}. Planet d spends part of its orbit in the HZ. Due to the non-zero orbital eccentricity, $T_{eq}$ is a function of time (or phase angle) in this case. Thus, the $T_{eq}^{f=4}$ for planet d varies from 229~K at periastron to 154~K at apastron, causing it to periodically dip into and out of the HZ.

\subsection{GJ~15A}
\label{chapter:systems:sec:gj15a}

GJ~15A (HD~1326A) is another very nearby \citep[distance 3.6pc;][]{jao05,hen06} late-type dwarf with a low-mass planet ($M$ sin$i = 5.35$ \mearth) in a 11.4-day orbit \citep{how14}. The star itself is the more massive component of a stellar binary system with a separation of about 150 AU and a period of 2600 years. The system's proximity and brightness make it a very attractive target for various follow-up studies. 

Based on CHARA interferometry and consistent with previously obtained, lower-precision measurements of GJ~15A's diameter, the stellar radius is calculated to be $(0.388 \pm 0.002)$ \rsun, the effective temperature \teff = (3602 $\pm$ 14) K, and the luminosity (0.0228 $\pm$ 0.0003) \lsun  (\citealt{how14}; see also Table \ref{tab:systems:dwarfs}).


\subsection{GJ~876}
\label{chapter:systems:sec:gj876}

The nearby (4.7 pc), late-type (M4), multiplanet host GJ~876 (HIP~113020) has at least four non-transiting planets in orbit \citep{cor10,riv10}. The inclination angle of the system is known to be $59.5^{\circ}$. The planet masses range from 6.83 $M_{Earth}$ to 2.28 $M_{Jup}$ in orbital distances that range from 0.02 to 0.33 AU with periods between 1.94 days and 124.26 days \citep{riv10}. CHARA interferometry was used to determine stellar radius = $(0.376 \pm 0.006)$ \rsun,  $T_{\rm eff}$ = (3131 $\pm$ 19) K, and $L$ = (0.0122 $\pm$ 0.0002) \lsun  (\citealt{von14}; see also Table \ref{tab:systems:dwarfs}).

\begin{figure}										
\centering
\includegraphics[width=\linewidth,angle=270]{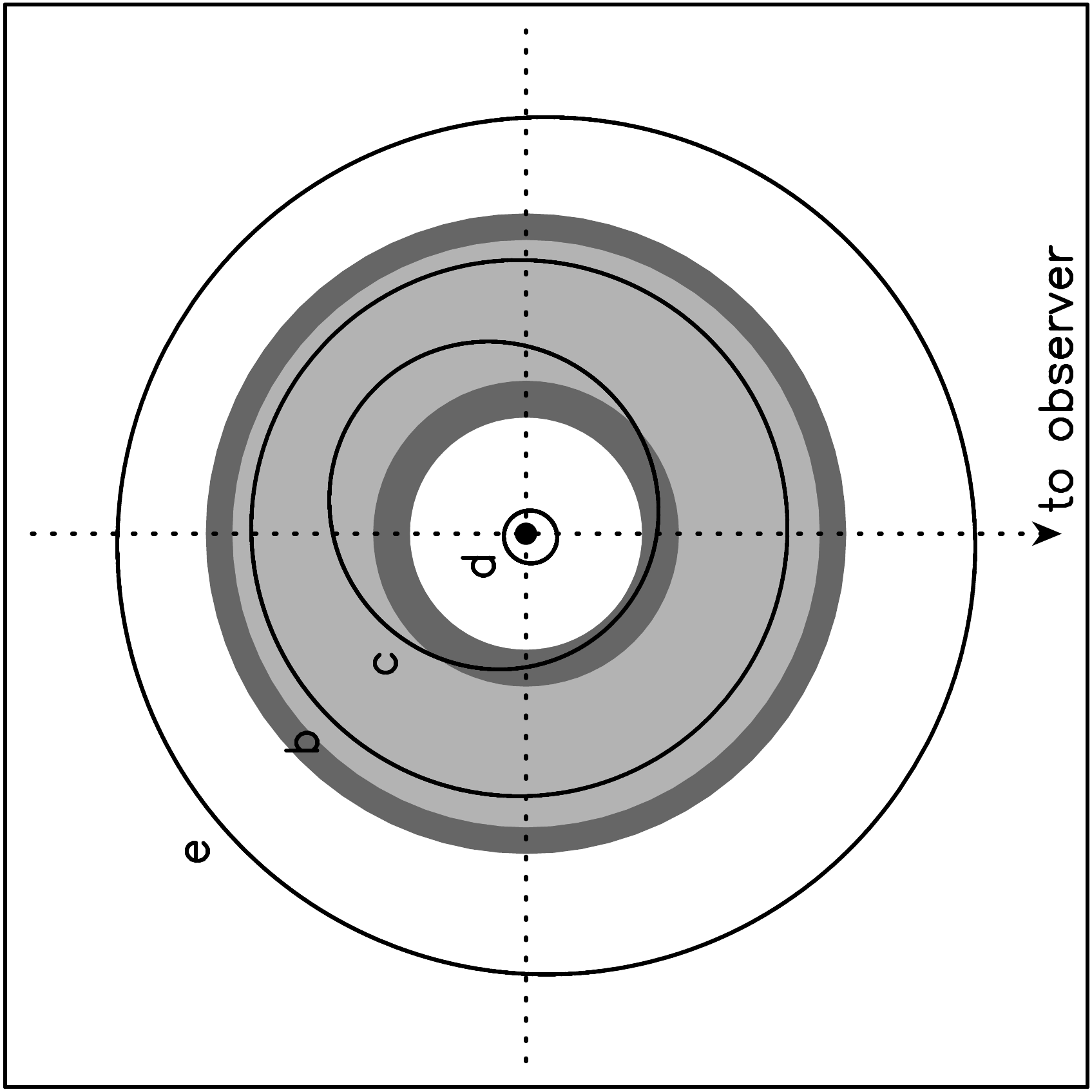}
\caption{Architecture of the GJ~876 system. The conservative HZ is shown in light grey, the optimistic HZ includes the light grey and dark grey regions. Planets b and c spend their entire orbits in the optimistic HZ. Planet b spends its entire orbit in the conservative HZ, whereas planet c spends 68.5\% of its orbital period in it. For details, see \S \ref{chapter:systems:sec:gj876}, and Table \ref{tab:systems:dwarfs}. Orbital parameters for the planets are from \citet{riv10}. For scale: the size of the box is 0.8 AU $\times$ 0.8 AU. Figure adapted from \citet{von14}.}
\label{fig:gj876_HZ}
\end{figure}

These values imply optimistic / conservative HZ boundaries of 0.09--0.24 AU / 0.12--0.23 AU. For an image of the system architecture, see Fig. \ref{fig:gj876_HZ} in which the conservative HZ is shaded in light grey and the optimistic HZ additionally containst the dark grey parts. Planet b spends its entire orbit in the conservative HZ, whereas planet c spends 68.5\% of its orbital period in it. Both planets spend their entire orbits in the optimistic HZ. The size of the box is 0.8 AU $\times$ 0.8 AU. 

Analogous to Section \ref{chapter:systems:sec:gj581}, we can calculate the equilibrium temperatures $T_{eq}$ for the GJ~876 planets. If we assume a Bond albedo value of 0.3, the planetary equilibrium temperatures for $f$ = 4 are 587 K (planet d), 235 K (planet c), 186 K (planet b), and 147 K (planet e). The $T_{eq}$ values for $f$ = 2 are 698 K (d), 280 K (c), 221 K (b), and 174 K (e). These values scale as $(1-A)^\frac{1}{4}$ for other Bond albedo values (Equation \ref{eq:equitemp}). 

Previously published values for the stellar radius of GJ~876, based on indirect methods, are significantly below our directly determined value: $0.24 R_{\odot}$ \citep{zak79} and $0.3 R_{\odot}$ \citep{lau05,riv10}. $0.3 R_{\odot}$ is the value that is frequently used in the exoplanet literature on GJ~876 -- note that that value is 25\% lower and more than 10$\sigma$ below the directly determined value in Table \ref{tab:systems:dwarfs}, providing additional incentive to help solve the ongoing discrepancy between theoretically predicted and empirically determined stellar radii particularly for late-type dwarfs. For the system effective temperature, literature values feature a seemingly bimodal distribution of values: $3130$~K \citep{dod11}, $3165 \pm 50$~K \citep{hou12}, 3172~K \citep{jen09}, $3765^{+477}_{-650}$~K \citep{amm06}, and 3787~K \citep{but06}.


\subsection{61~Vir}
\label{chapter:systems:sec:61vir}

61~Vir (HD~64924) is a nearby (8.6 pc), G7 dwarf with three planets in orbit. Their periods range from 4.2 to 124 days and minimum masses from 5.1 to 24 $M_{Earth}$ \citep{vog10b}. CHARA interferometry produced a stellar radius $(0.987 \pm 0.005)$ \rsun, an effective temperature of (5538 $\pm$ 13) K, and a luminosity of (0.825 $\pm$ 0.003) \lsun  (\citealt{von14}; see Table \ref{tab:systems:dwarfs}). All of the exoplanets reside closer to the parent star than the inner edge of the system HZ. 
 
Previously published values for 61~Vir's stellar radius are consistent with \citet{tak07a} or $\sim 2 \sigma$ below \citep{val05} the value quoted here. Effective temperatures in the literature are generally in agreement with the value in Table \ref{tab:systems:dwarfs} \citep{val05,ecu06}.

In contrast to the late-type dwarfs that have not appreciably evolved since their births, 61~Vir's stellar parameters can be used as input to isochrone fitting to estimate stellar mass and age. \citet{von14} thus calculate an age of $8.6 \pm 5$~Gyr and mass of $0.93 \pm 0.05 M_{\odot}$.


\subsection{HD~69830}
\label{chapter:systems:sec:hd69830}

HD~69830 is a nearby (12.5 pc), K0 dwarf with three Neptune-mass exoplanets (minimum masses between 10 and 18 $M_{Earth}$) in slightly eccentric orbits with periods between 8.7 and 197 days. A very interesting feature of this system is that it has a prominent asteroid belt initially discovered via a Spitzer 24$\mu$m excess \citep{bei05} at a distance of 0.5--1 AU from the star. CHARA interferometry enables the measurement of the stellar radius $(0.9058 \pm 0.0190)$ \rsun, stellar $T_{\rm eff}$ (5394 $\pm$ 62) K, and stellar luminosity (0.622 $\pm$ 0.014) \lsun  (\citealt{tan15}; see Table \ref{tab:systems:dwarfs}). 

The stellar parameters allow for the calculation of the optimistic and conservative system HZ (0.605--1.442 AU and 0.767--1.368 AU, respectively). The outermost planet (d) spends the majority of its elliptical orbit in the optimistic HZ. Isochrone fitting with the use of echelle spectra yields a system age of 7.5 $\pm$ 3 Gyr, implying that the asteroid belt is not primordial but needs to either be replenished through asteroid impacts or be a short-lived phenomenon \citep{tan15}.


\subsection{HR~8799}
\label{chapter:systems:sec:hr8799}

HR~8799 (HD~218396) is a young, chemically peculiar, early-type pulsator star that exhibits an excess of infrared radiation longwards of 20$\mu$m, indicative of thermal emission from a debris disk \citep{su09}. It also is known to host four directly imaged substellar-mass companions \citep{mar08,mar10}. The age of the system is instrumental in determining the companion masses, putting them either into the exoplanet mass range for very young ages (10 -- 100 Myrs) or into the brown dwarfs domain for older ages. Age estimates for this system, based on various methods, vary widely: they range from 20 Myrs to more than 1.6 Gyrs -- see the discussion in the introduction of \citet{bai12}, in particular their table 1. This discrepancy in calculated system age motivated the \citet{bai12} interferometric study of this star using CHARA, whose results, when coupled with trigonometric distance and bolometric flux measurement, impose model-independent constraints on the calculation of stellar mass and age. 

HR~8799's radius, effective temperature, and luminosity are measured to be $(1.45 \pm 0.05) R_{\odot}$, (7163 $\pm$ 84) K, and ($4.973 \pm 0.275) L_{\odot}$ \citep{bai12,boy13}, respectively. \citet{bai12} note that the radius is 8\% larger than the one used in a number of aforementioned studies to determine the system age. Following the arguments that the chemical pecularity of the system is caused by the debris disk and the stellar metallicity is, in fact, close to solar, \citet{bai12} calculate the stellar mass to be around 1.5 solar masses, and the stellar age to be young ($\leq 0.1$ Gyr), which is supported by dynamical arguments. If the abundance of HR~8799 is indeed close to solar, then that implies that the companion are of planetary mass, i.e., below the brown dwarf boundary. 



\subsection{55 Cancri}
\label{chapter:systems:sec:55cnc}

55 Cancri (HD~75732; $\rho$ Cancri; 55~Cnc) is a nearby (12.5 pc) G/K dwarf currently known to host five extrasolar planets with periods between around 0.7 days and 14 years and minimum masses between 0.026 and 3.84 $M_{Jup}$ \citep{but97,mar02, mca04, fis08, daw10}. The innermost planet (e) transits its host star \citep{dem11, win11}. 55~Cnc is a very attractive target for interferometric studies to directly determine or constrain stellar and planetary system parameters for the following reasons: (1) the close proximity of the system enables the interferometric measurement of the stellar diameter, (2) the range of the planets' orbital periods and semi-major axes puts some of them into or close to the system HZ, and (3) the transiting planet whose diameter is a direct function of the stellar diameter. 


\subsubsection{Stellar Parameters}
\label{chapter:systems:sec:55cnc:parameters}

The CHARA interferometric studies presented in \citet{von11b} and \citet{lig16} measure 55~Cnc's limb-darkening corrected angular diameters whose weighted average, when combined with SED fitting \citep{von11b,von11a,boy13,lig16} and trigonometric parallax measurements \citep{van07}, produce the following system parameters: stellar diameter $R = (0.945 \pm 0.010) R_{\rm \odot}$, $T_{\rm EFF} = (5178 \pm 17)$ K, and $L = (0.578 \pm 0.011) L_{\odot}$. The value for $\theta_{\rm LD}$ in Table \ref{tab:systems:dwarfs} is consistent with the previously published value in \citet{van09} at the 1.5$\sigma$ level. Furthermore, it exactly corresponds to the angular diameter required for 55~Cnc to fall onto the $T_{\rm EFF}$ versus $(V-K)_0$ relation in \citet{van09}; see their equation 2 and section 5.4.1. It is also worth noting that the \cite{von11b} and \citet{lig16} numbers for 55~Cnc's stellar radius are consistent with the calculated value in \citet{fis08} based on stellar parameters in \citet{val05}.

When comparing the astrophysical parameters for 55 Cnc's stellar luminosity and effective temperature to the Yonsei-Yale stellar isochrones \citep{dem04,kim02, yi2001} with [Fe/H]~$= 0.31$ \citep{val05,fis08}, \citet{von11b} obtain a stellar age of 55~Cnc of 10.2 $\pm$ 2.5~Gyr and stellar mass of 0.905 $\pm$ 0.015$M_{\odot}$, which are in good agreement with the values published in \citet{val05} and \citet{fis08}, based on spectroscopic analyses, and \citet{wri04} and \citet{mam08}, based on the Ca II chromospheric activity indicators and gyrochronology relations. Finally, the surface gravity $\log g = 4.45 \pm 0.01$ is readily computed from radius measurement and mass estimate.


\subsubsection{Habitable Zone and Planet 55 Cnc f}
\label{chapter:systems:sec:55cnc:planets}

\begin{figure}										
\centering
\epsfig{file=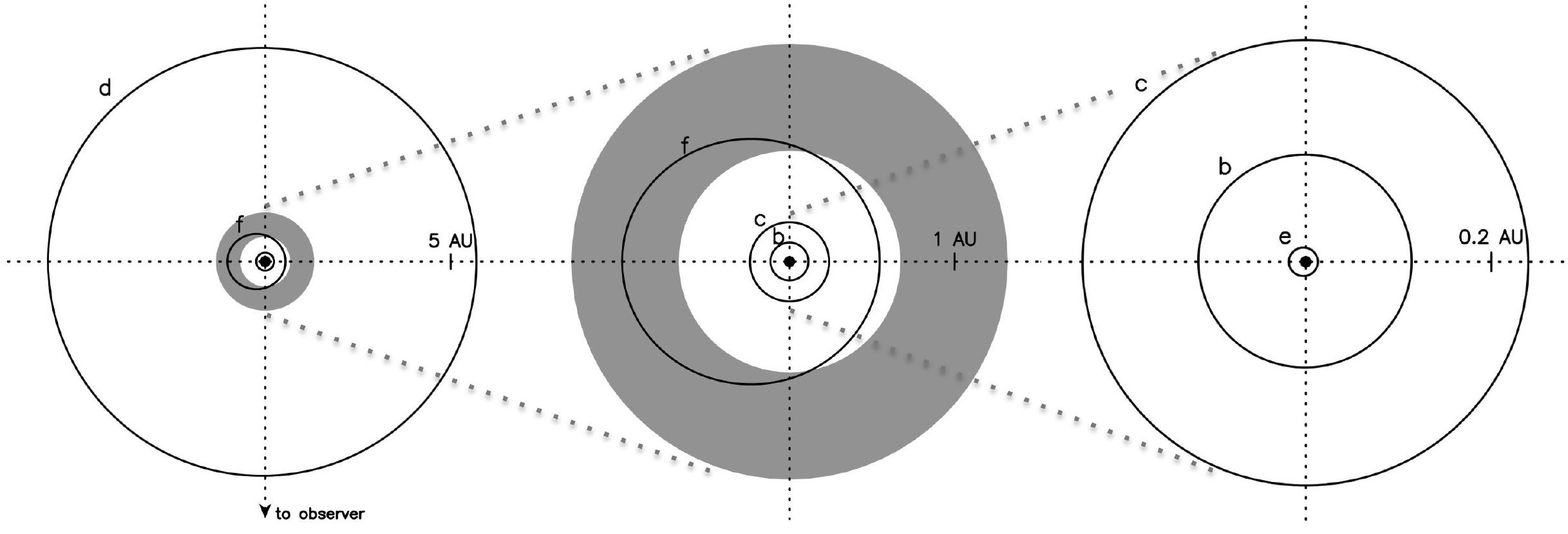,width=\linewidth,,clip=} \\
  \caption{Orbital architecture of the 55~Cnc system, centered on the star at increasing zoom levels indicated by the dashed lines from left to right (note the different scales on the ordinate). The habitable zone is indicated by the gray shaded region. Orbital element values are from \citet{daw10}. Planet f periodically dips into and out of the HZ during its elliptical ($e \simeq 0.3$) orbit (left and middle panels). Planets b, c, e (the transiting super-Earth), and d are not in the system HZ. For details, see \S \ref{chapter:systems:sec:55cnc:planets}. Figure adapted from \citet{von11b}.}
  \label{fig:fig4}
\end{figure}

\citet{von11b} use the equations in \citet{und03} \& \citet{jon10} to calculate inner (0.67~AU) and outer (1.32~AU) edges of the HZ from stellar luminosity and effective temperature. The HZ is shown as the gray-shaded region in Figure \ref{fig:fig4}, which illustrates the architecture of the 55~Cnc system at different spatial scales\footnote{The distinction between optimistic and conservative HZ boundaries did not exist in 2011 (see \S \ref{chapter:parameters:sec:hz}).}. 

Equilibrium temperatures $T_{eq}$ for the five planets can be calculated using Equation \ref{eq:equitemp}. All of 55~Cnc's known planets, except planet f, are either located well inside or beyond the system's HZ. Figure \ref{fig:fig4} shows the orbital architecture of the 55~Cnc system, centered on the star at increasing zoom levels. The habitable zone is indicated by the gray shaded region. The inner planets b, c, e (the transiting super-Earth; see \S \ref{chapter:systems:sec:55cnc:planet_e}), and the outer planet d are not in the system HZ. 55~Cnc f \citep[$M \sin i = 0.155  M_{Jup} = 49.3 M_{Earth}$; table 10 in][]{daw10}, however, is in an elliptical orbit ($e \simeq 0.3$) during which it spends about 74\% of its $\sim$ 260 days orbital period inside the system HZ at a time-averaged distance of 0.82 AU from the parent star \citep{von11c}. Thus, $T_{eq}$ for planet f is a function of  phase angle. For the $f=2$ scenario (no heat redistribution between day and night sides) and a Bond albedo of $A=0.29$, planet f's time-averaged dayside temperature is 294 K, and varies between 263 K (apastron) and 359 K (periastron). For the scenario of even heat redistribution ($f=4$) and $A=0.29$, planet f's time-averaged surface temperature $T_{eq}^{f=4} = 247$ K, with a variation of 221~K at apastron to 302~K at periastron. 

Planet f's long-period, elliptical orbit makes any kind of tidal synchronization unlikely. Further taking into account its mass that is typical of that of gas giant planets, $f=4$ appears to be a much more likely scenario than $f=2$ or similar. For the above calculations of $T_{eq}$, an orbital eccentricity of $e_f$ = 0.3 was assumed; however, \citet{daw10} state that the orbital eccentricity is most likely between 0.13 and 3. The differences between the temperatures at apastron and periastron calculated above, as well as the planet average $T_{eq}$, decrease if $e_f$ is smaller than 0.3 \citep{von11b}.  

55~Cnc f is likely too massive to harbor liquid water on any planetary surface since it probably does not have a defined surface \citep{skl07}.  In terms of an actual {\it habitable object} in the 55~Cnc system, there are thus two potential candidates: a massive moon in orbit around planet f or an additional low-mass planet in or near the HZ. 

Could 55~Cnc f host a potentially habitable moon? The short answer is yes, but the moon would have to be massive enough to retain a sufficiently thick atmosphere \citep{wil97,agn06}, or the true orbital eccentricity of planet f would have to be lower that 0.3 to decrease the equilibrium temperature of the hypothetical planet-moon system. 

Could there be an unseen, low-mass planet with a stable orbit in the system HZ? The short answer to that is also yes. Based on dynamical studies, there are small ranges in distance for which orbits of low-mass planets are stable for $e_f\sim 0.3$: either the outermost regions of the HZ or a small torus between 1.02-1.04 AU at the exterior 3:2 mean motion resonance with planet f \citep{ray08}. For lower values of $e_f$, these ranges increase in size. 


\subsubsection{Transiting Planet 55~Cnc e}
\label{chapter:systems:sec:55cnc:planet_e}

Any calculation of planetary radius of a transiting planet of the form 

\begin{equation}\label{eq:fluxdecrement}
  \delta flux = (\frac{R_p}{R_\star})^2,
\end{equation}

where $\delta flux$ implies the measured flux decrement during planet transit, represents a lower limit to the planetary radius because it assumes that the planet is an opaque spot superimposed onto the stellar disk during transit. In reality, the temperature contrast between star and planet also enters the equation \citep{win10}. However, the close proximity of 55~Cnc e to its parent star makes any atmosphere retention extremely unlikely, rendering the above assumption adequate. The interferometric study yields the stellar radius, and the flux decrement is produced by the two concurrent studies that announced the transit signature of planet e in 2011.  

\citet{win11} obtain $\frac{R_p}{R_\star}$ = $0.0195 \pm 0.0013$ and $M_{p}$ = $8.63 \pm 0.35 M_{\oplus}$ implying $R_{p} = 2.007 \pm 0.136 R_{\oplus}$ and a planetary bulk density of $5.882 \pm 0.728$ g cm$^{-3}$ or $1.067 \pm 0.132 \rho_{\oplus}$.
\citet{dem11} measure $\frac{R_p}{R_\star} = 0.0213 \pm 0.0014$ and assume a planetary mass of $7.98 \pm 0.69 M_{\oplus}$, which produces planetary radius of $2.193 \pm 0.146 R_{\oplus}$ and a bulk density of $4.173 \pm 0.602$ g cm$^{-3}$, corresponding to $0.757 \pm 0.109 \rho_{\oplus}$.
For more information, see figure 3 in \citet{win11} and figures 5 \& 6 in \citet{dem11}. It should be noted that, although 55~Cnc is one of the brightest known stars with a transiting planet, the amplitude of the transit signal is very small, making the determination of planet radius very difficult.


\subsection{GJ~436}
\label{chapter:systems:sec:gj436}

GJ~436 (HIP~57087) is a well-studied, nearby (10.2 pc) M3 dwarf that is known to host a transiting, Neptune-sized exoplanet in a 2.64-day orbit \citep{but04,demory07,deming07,gil07b,gil07a,cac09,bea08,pon09,fig09,bal10b,sou10b}. This system is a particularly good target for interferometry studies due to the following aspects: 

\begin{itemize}

\item GJ~436 is an M dwarf. The obvious assumption in the calculation of planetary radius and density is the knowledge of the stellar radius calculated from, for instance, stellar models. 
For spectral types around M3V, the well-documented offset between model radii and directly measured counterparts is on the order of a few to ten percent \citep[][and references therein]{tor07,mercedes2007,lop07,von08,boy10,man13,man15b}; but see also \citet{dem09}. 
\item The transit depth is much larger than that of, e.g., 55~Cnc e (\S \ref{chapter:systems:sec:55cnc:planet_e}), making very precise measurements much more straightforward. 
\item The planet-to-stellar mass ratio is greater than for 55~Cnc e as well, improving the precision of mass determination based on radial velocity studies. 
\item There are a number of space- and ground-based studies of this system whose results can be combined with interferometry results to fully characterize both the stellar and planetary parameters. 

\end{itemize}


\subsubsection{Directly Determined Stellar Parameters}
\label{chapter:systems:sec:gj436:parameters}

CHARA interferometry presented in \citet{von12}, coupled with trigonometric parallax values from \citet{van07} and SED fitting from \citet{man13}, produces the following directly determined stellar parameters: $R_{\star} = (0.455 \pm 0.018) R_{\rm \odot}$\footnote{Especially for a transit depth much greater than the one of 55~Cnc e, the question can be asked whether an interferometric measurement obtained during planetary transit or during the presence of star spots could produce a radius estimate that is thus artificially reduced. As seen in \citet{van08a}, this effect is expected to be very small. Interferometric visibility variations in a single data point due to a transiting planet will impact current diameter measurements at the $\delta \theta \simeq 0.6$\% level, buried in the measurement noise of the result based on all obtained visibility measurements. Since spots feature lower contrast ratios than transiting planets, expected visibility variations are even smaller.}, $T_{\rm EFF} = (3505 \pm 55)$ K\footnote{Due to the grazing  transit in the GJ~436 system, limb-darkening models play a larger role in the calculation of the planetary radius calculated than for central transits. Knowing the stellar effective temperature to a higher precision thus provides particularly important constraints for this system.}, and $L = (0.0281 \pm 0.0014) L_{\odot}$ (see Table \ref{tab:systems:dwarfs}).

A comparison between the interferometrically determined radius and that predicted by stellar models -- the \citet{bar98} models with 5 Gyr age, solar metallicity, 0.44 $M_{\rm \odot}$ mass for GJ~436 \citep{man07,tor07} -- readily confirms the radius discrepancy for M dwarfs mentioned above: the interferometry radius ($0.455 R_{\rm \odot}$) exceeds the theoretical one ($0.409 R_{\rm \odot}$) by 11\%. This aspect is further discussed in \citet{tor07}, where the radius of GJ~436 is found to be overly inflated for its mass. 

If, however, light curve analysis of the transiting planet is used in conjunction with the the stellar mass values from \citet{man07} or \citet{tor07}, literature values of the stellar radius of GJ~436 are in agreement with the direct \citet{von12} measurement \citep{gil07a,gil07b,deming07,shp09,bal10a,sou10b,knu11}, clearly illustrating the value of transiting planets for not just planetary but also stellar physics. See \citet{tor07} for a very good review and more details on this aspect. 

In the same manner as for GJ~581 (\S \ref{chapter:systems:sec:gj581}) and 55~Cnc (\S \ref{chapter:systems:sec:55cnc}), the system HZ can be calculated to be located at 0.16 -- 0.31 astronomical units (AU) from GJ~436, clearly beyond the orbital semimajor axis of GJ~436b ($a \simeq 0.03$ AU).


\subsubsection{Calculated Parameters for Star and Planet}
\label{chapter:systems:sec:gj436:mcmc}

The knowledge of stellar radius provides for a way to calculate, rather than assume, a stellar mass, instead of the other way around. The study by \citet{von12} uses the directly measured stellar parameters (see \S \ref{chapter:systems:sec:gj436:parameters} and Table \ref{tab:systems:dwarfs}) and combines them with a global analysis of literature time-series photometry and radial velocity (RV) data to obtain a characterization of the system as a whole, including stellar and planetary physical and orbital parameters.   

In a transiting exoplanet system, the shape of the light curve depends in part on the mean stellar density \citep[e.g.,][]{sea03,soz07,tin11}. Consequently, \citet{von12} follow the technique described in \citet{cameron2007} and combine publicly available RV and photometry time-series data on GJ~436 in a global Markov Chain Monte Carlo (MCMC) analysis, setting the directly determined stellar parameters to be fixed. The derived astrophysical parameters fully characterize the system (planet and star) and include planetary and stellar masses, bulk densities, radii, surface gravities, plus orbital elements, along with associated uncertainties \citep[see tables 2 and 4 in][]{von12}.


\subsection{HD~209458}
\label{chapter:systems:sec:hd209458}

HD~209458 was the first known transiting planet \citep{cha00} and has been studied very extensively over the course of the past two decades. The system consists of a solar-type host star at a distance of about 47 pc with an inflated hot Jupiter in a 3.5-day orbit. There are myriad different data products for the star and exoplanet, including very-high-precision photometry and spectroscopy time series. HD~209458 is the first system for which spectral lines for both star and planet were individually detected \citep{sne10}, allowing for the independent determination of component masses since the inclination angle of the planetary orbit with respect to the line of sight is known. 

The study presented in \citet{boy15} uses optical CHARA interferometry and flux-calibrated spectrophotometry data to determine HD~209458's stellar parameters to be $R_{\star}=$(\radtwo) $R_{\rm \odot}$, $T_{\rm eff} =$(\tefftwo) $K$, and $L=$(\lumintwo) $L_{\odot}$ (see Table \ref{tab:systems:dwarfs}). The angular diameter of HD~209458 is just slightly over 0.2 mas, which is at the current resolution limit of the CHARA Array. Coupled with space-based, infrared transit depth measurements \citep{bea10} and planetary and stellar mass values from \citet{sne10}, \citet{boy15} obtain empirically determined values for the planetary radius of (\radtwop) $R_{Jup}$ and the planetary bulk density of \rhotwocgsp, which is similar to the density of cork. Other directly determined parameters include stellar bulk density as well as the surface gravities for both planet and parent star (see table 3 in \citealt{boy15}).

Previously published works on HD~209458's stellar parameters, e.g., \citet{cod02}, \citet{tor08}, and \citet{sou10,sou11}, agree well with the value in Table \ref{tab:systems:dwarfs}. Furthermore, the surface brightness relations published in \citet{boy14}, which are empirical relations between broad-band colors and angular stellar diameter (see \S \ref{chapter:future:sec:methods}), accurately predict HD~209458's radius.


\subsection{HD~189733}
\label{chapter:systems:sec:hd189733}

The aforementioned study that presented interferometry results on HD~209458 (\S \ref{chapter:systems:sec:hd209458}) used the same methods to investigate the transiting planet host HD~189733, an early K dwarf at a distance of 19.2 pc \citep{boy15}: $R_{\star}=$(\radone) $R_{\rm \odot}$, $T_{\rm eff} =$(\teffone) $K$, and $L=$(\luminone) $L_{\odot}$ (see Table \ref{tab:systems:dwarfs}). Due to its brightness, HD~189733 has been similarly well studied as HD~209458, resulting in availability of many different data products. As for HD~209458, spectral line detection for both parent star and transiting exoplanet in the HD~189733 system enables the calculations of both component masses individually \citep{dek13,rod13}. Using the \citet{dek13} component mass values and transit depth measurements from \citet{ago10} in combination with their interferometry and spectrophotometry data, \citet{boy15} empirically determine the 
planetary radius to be (\radonep) $R_{Jup}$ and the planetary bulk density of \rhoonecgsp, consistent with the density of butter. As for HD~209458, table 3 in \citet{boy15} contains the other directly determined parameters such as stellar bulk density and surface gravities for both planet and parent star.

The measured angular diameter from \citet{boy15} agrees well with the predicted one from the surface brightness relations in \citet{boy14} and the one previously obtained via interferometry \citep{bai07a}, but the error estimate on the stellar diameter is better by a factor of about 3 in \citet{boy15} due to the fact that the observations were obtained at optical ($R$) instead of near-infrared ($H$) wavelengths (see \S \ref{chapter:parameters:sec:radius}). Unlike HD~209458, however, the measured stellar diameter is inconsistent with model predictions at the $> 2 \sigma$ level: the model predictions underestimate the stellar diameter for HD~189733 by 5-10\%, depending on the assumptions and/or models \citep{tor08}, confirming this previously reported discrepancy \citep{tor07,tor08,boy12b,von12,spa13}; see figure 3 and section 5 in \citet{boy15}. 

In order to adjust stellar models to comply with observational data in the literature, especially the interferometrically determined radius, \citet{boy15} vary a number of different model input parameters, such as system age, different aspects of stellar composition, e.g., metalliticy, primordial helium abundance, etc., plus magnetic fields and starspots, and convection. The only scenario that harmonizes model output with all available literature observables consists of lowering of the numerical value of the solar-calibrated mixing-length parameter $\alpha_{\rm MLT}$, which is a measure of the distance in units of the  pressure scale height over which a parcel of gas dissipates its energy and adjusts its temperature to its surroundings. 

This finding provides motivation to re-calibrate mixing-length parameters for stars with lower masses than the sun in order to reduce or even eliminate the consistent discrepancy between directly determined stellar radii and effective temperatures vs. the ones predicted by stellar models. 


\subsection{HD~219134}
\label{chapter:systems:sec:hd219134}

HD~219134 (GJ~892) is a nearby (6.5 pc), late-type dwarf with six or more planets in tightly packed orbit \citep{vog15}. \citet{mot15} used {\it Spitzer} to detect a transit signature of the innermost planet, a super-earth in a 3-day orbit. 

The host star was studied in \citet{boy12b,man13}: $R_{\star} = (0.778 \pm 0.005) R_{\rm \odot}$, $T_{\rm eff} = (4773 \pm 22)$ K, and $L = (0.283 \pm 0.004) L_{\odot}$ (Table \ref{tab:systems:dwarfs}). \citet{hub16} furthermore presents preliminary results on radius determination via optical interferometry and demonstrates excellent agreeement between the \citet{boy12b} NIR and his optical CHARA data. 

Knowledge of the stellar radius coupled with the {\it Spitzer} flux decrement during transit produces the planetary radius of 1.6 \rearth\ and a planetary density similar to the one of Earth \citep{mot15}. The system's proximity and brightness makes it well suited for follow-up studies. 


\subsection{Other Systems}
\label{chapter:systems:sec:others}

In addition to the systems pointed out individually in the Sections above, there are obviously a number of other ones that are interesting in their own right. A few examples of those are briefly summarized below. System parameters are given in Table \ref{tab:systems:dwarfs}.

{\bf HD~9826 ($\upsilon$ And)} represents the first discovery of a multiple planet system around a main sequence star and the first multiple planet system found in a binary star system. The planets are furthermore believed to be in non-coplanar orbits. The exoplanet host star was characterized via interferometry and SED fitting in \citet{bai08,lig12,boy13,lig16}. 

The {\bf Alpha Centauri} triple star system is our closest stellar neighbor and hosts a planet around the faintest component. The brighter two stars were characterized in \citet{ker03a,big06,boy13,ker16,ker17}.

{\bf Fomalhaut} is one of very few stars with a directly imaged substellar companion in orbit. Interferometry on Fomalhaut was performed in \citet{dif04,dav05}.

{\bf HD~33564} is a bright (naked-eye) F-star with a substellar companion in the system HZ. The host star was characterized in \citet{von14}. 

Finally, the TERMS survey \citep{kan09} aims to characterize exoplanetary systems with respect to their orbital and astrophysical properties and looks for transit signatures of known RV planets. Knowledge of the host star radii is important to determine planetary radii if transits are detected and to constrain statements regarding planetary radii if no transits are detected at a given relative photometry precision, e.g., for the targets {\bf HD~38529} \citep{hen13} and {\bf 70~Vir} \citep{kan15}.


\section{Summary of Results}
\label{chapter:systems:sec:summary}

The most important of the study presented here and the ones on which it reports is the accumulation of the individual results into a sizeable number of stars with directly determined diameters that cover a wide range in radii and effective temperatures (Table \ref{tab:systems:dwarfs} and Figure \ref{fig:HRD_EHS_blue}). Various subsets of these insights have been and are being used to calibrate semi-empirical relations between observables and the predictions for astrophysical parameters of stars too faint or too distant to study interferometrically, which obviously constitutes the vast majority of stars in the Milky Way. The same applies to providing constraints for the continuous improvements of stellar models.

Another insight from Table \ref{tab:systems:dwarfs} is that we confirm the discrepancy between predicted and measured stellar radii, which increases for late-type dwarfs or stars much hotter than the sun -- canonical values here are 5--10\% in diameter and 3--5\% in effective temperature, but with examples that exceed these values. Reasons for this discrepancy have been proposed in the past as being stellar metallicity, activity, etc. We find for one very well studied system (HD~189733) that it may instead be the convective mixing-length parameter $\alpha_{\rm MLT}$ (\S \ref{chapter:systems:sec:hd189733}). 

Figure \ref{fig:HRD_EHS_blue} furthermore indicates that there is no systematic difference in terms of astrophysical parameters between planet-hosting stars and stars not currently known to host planets. This is no surprise, of course, since the designation of ``stars not currently known to host planets" may not indicate that they do not have (thus far undiscovered or unpublished) planets in orbit. One such example is HD~219134 whose parameters were published in \citep{boy12b} and \citep{man13}, but whose planets were not known to the astronomical community until 2015 \citep{vog15,mot15}. In addition, if there were a significant difference between stars with and without planets, the origins of this difference would most likely have to lie in the formation mechanisms, which, given the insignificant fraction of system mass contained in planets, would appear surprising. 

Finally, we generally find very good agreement for different angular diameter measurements across different studies even when different interferometers were used. This agreement has been documented before \citep[e.g.,][]{boy12b,boy13,hub16}.


%
%
%
\chapter{Future Work}
\label{chapter:future} 




\section{The Possible: Future Targets}
\label{chapter:future:sec:targets}

CHARA and VLTI continue to the be the currently most productive interferometers in the world for the measurements of stellar diameters, with the Naval Precision Interferometer (NPOI) contributing to the measurements of diameters of larger-diameter, brighter stars. NPOI will undergo substantial upgrades to its infrastructure in the next few years and thus open up parameter space currently unavailable to other interferometers. Furthermore, upgrades to CHARA, in particular the installation of an adaptive optics system, in the near future will push CHARA's limits in both sensitivity and resolution to higher levels. 

Even with current capabilities\footnote{As of the time of this writing, CHARA can, in good observing conditions, observe stars as far south as around --30 degrees in declination, stars as faint as $V \simeq 11$ coupled with either $R \simeq 8$ for optical work or $H \simeq 8$ for NIR work, as well as stellar angular diameters as small as around 0.4 mas in the NIR and 0.25 mas in the optical. These numbers are (1) dependent on observing conditions, and (2) continually improving with ongoing system optimization and upgrades.}, however, the number of stars that can be characterized with CHARA is high, both with and without known exoplanets around them. The continuation of the surveys described in this work by various groups will continue to provide direct measurements of stellar astrophysical parameters and thus characterization of individual systems and the addition of constraints to stellar models and correlations between measurable quantities and non-observable astrophysical quantities. 

Of particular interest among the observable stellar systems are host stars with multiple planets in orbit and/or planet(s) in the respective HZs, transiting planets hosts, and late-type stars, irrespective of whether they host planets or not. As we showed above, interferometric measurements are especially useful for transiting planet systems (see Sections \ref{chapter:systems:sec:55cnc}, \ref{chapter:systems:sec:gj436}, \ref{chapter:systems:sec:hd209458}, \ref{chapter:systems:sec:hd189733}, and \ref{chapter:systems:sec:hd219134}). We are currently targeting one of the few currently known transiting system attainable with CHARA's capabilities: HD~97658. Its planet was first announced in \citet{how11} with a subsequent transit discovery using MOST in \citet{dra13} and follow-up study with {\it Spitzer} in \citet{van14}, plus a study of its atmosphere in \citet{knu14}. The planet is a super-earth (7.5 \mearth, 2.2 \rearth) in a 9.5 day orbit, and its planetary parameters can be much more tightly constrained with a direct radius measurement \citep[e.g., ][]{van14,knu14}. 

To repeat the motivation of our interferometric survey, it is impossible to overstate the importance of ``understanding the parent stars''. With ongoing improvements in both sensitivity and angular resolution of NIR and optical interferometric data quality, we continue to provide firm, direct measurements of stellar radii and effective temperatures for exoplanet hosts and stars in the low-mass regime. 


\section{The Impossible: Indirect Methods, Limits, and Beyond}
\label{chapter:future:sec:methods}

The number of stars accessible to interferometry is limited by brightness and/or angular size. Thus, for most stars in the Milky Way, other methods need to be used to determine stellar diameters and other astrophysical parameters, such as for the very interesting transiting systems GJ~1214 and HD~149026. Interferometric results are often used to calibrate these methods, however. 

The full spectrum of these methods is beyond the scope of this publication \citep[but see, e.g.,][for some examples]{boy13}. However, we outlined in \S \ref{chapter:parameters:sec:why} alternate approaches to obtaining stellar diameter via the Stefan-Boltzmann Law by measuring bolometric flux and determining effective temperature via spectroscopic studies and/or using stellar models. This probably represents one of the most widely used approaches, particularly for late-type dwarfs for which interferometry is difficult due to the reasons mentioned above \citep[e.g.,][and references therein]{mui12,roj12,roj13b,roj13a,man13,man15b,new15,gai16}. Particular care needs to be taken to use improved literature broad-band filter zero points and response functions when measuring bolometric flux of target stars to avoid systematic effects and underestimate temperature uncertainties \citep{boh14,man15a}.

The accuracy of tabulated limb-darkening coefficients can also be examined via interferometric observations. For this to be possible, one's data need to cover the higher lobes of the interferometric visibility function, i.e., past the first zero in Figure \ref{fig:55Cnc_vis}. Such observations are more difficult to accomplish due to the lower fringe contrast in the higher lobes, and they require larger angular diameters of the targets and/or longer baselines and/or shorter wavelengths. One recent such example is published in \citet{ker17} for the $\alpha$ Centauri system using VLTI. Future observations of this kind, as the resolving power of interferometers increases, will further reduce systematic uncertainties in the conversion from uniform disk to limb-darkening corrected diameters (see \S\ref{chapter:parameters:sec:radius}). 

The large and increasing\footnote{Two sizeable data sets of interferometric observations of evolved stars that are nearing publication are described in \citet{bai17}, based on NPOI data, and \citet{van17}, based on data taken with the since retired Palomar Testbed Interferometer (PTI).} number of stars with directly determined astrophysical parameters allows for the establishment of relations of observable quantities with parameters such as effective temperature or angular diameters that are applicable for distant and/or faint stars. These relations are almost entirely direct in the sense that no stellar models or spectral modeling are necessary. 

\begin{figure}										
\centering
\includegraphics[width=\linewidth]{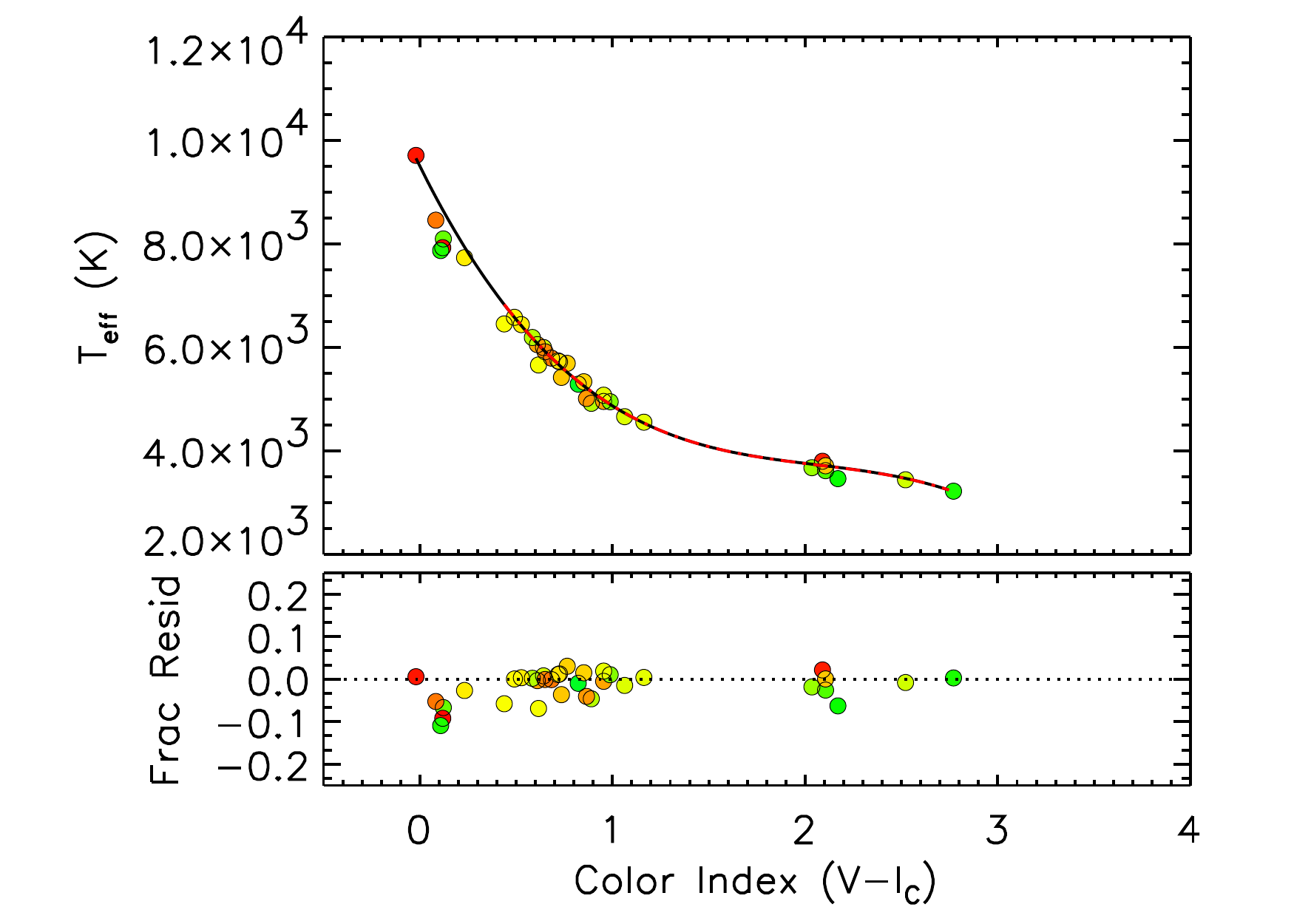}
\caption{Relation between stellar effective temperature of main sequence stars and broad-band $V-I_c$ color. The black line corresponds to the best fit, which is a 3rd order polynomial based on 34 stars fulfilling various criteria. The scatter of data around the fit is around 3\%. The different colors of the data points correspond to different metallicity values -- the scale is the same as shown in Fig. \ref{fig:theta_vs_V-Ic}. Figure adapted from \citet{boy13}. See \S \ref{chapter:future:sec:methods} for more details.}
\label{fig:Temp_vs_V-Ic}
\end{figure}

\begin{figure}										
\centering
\includegraphics[width=\linewidth]{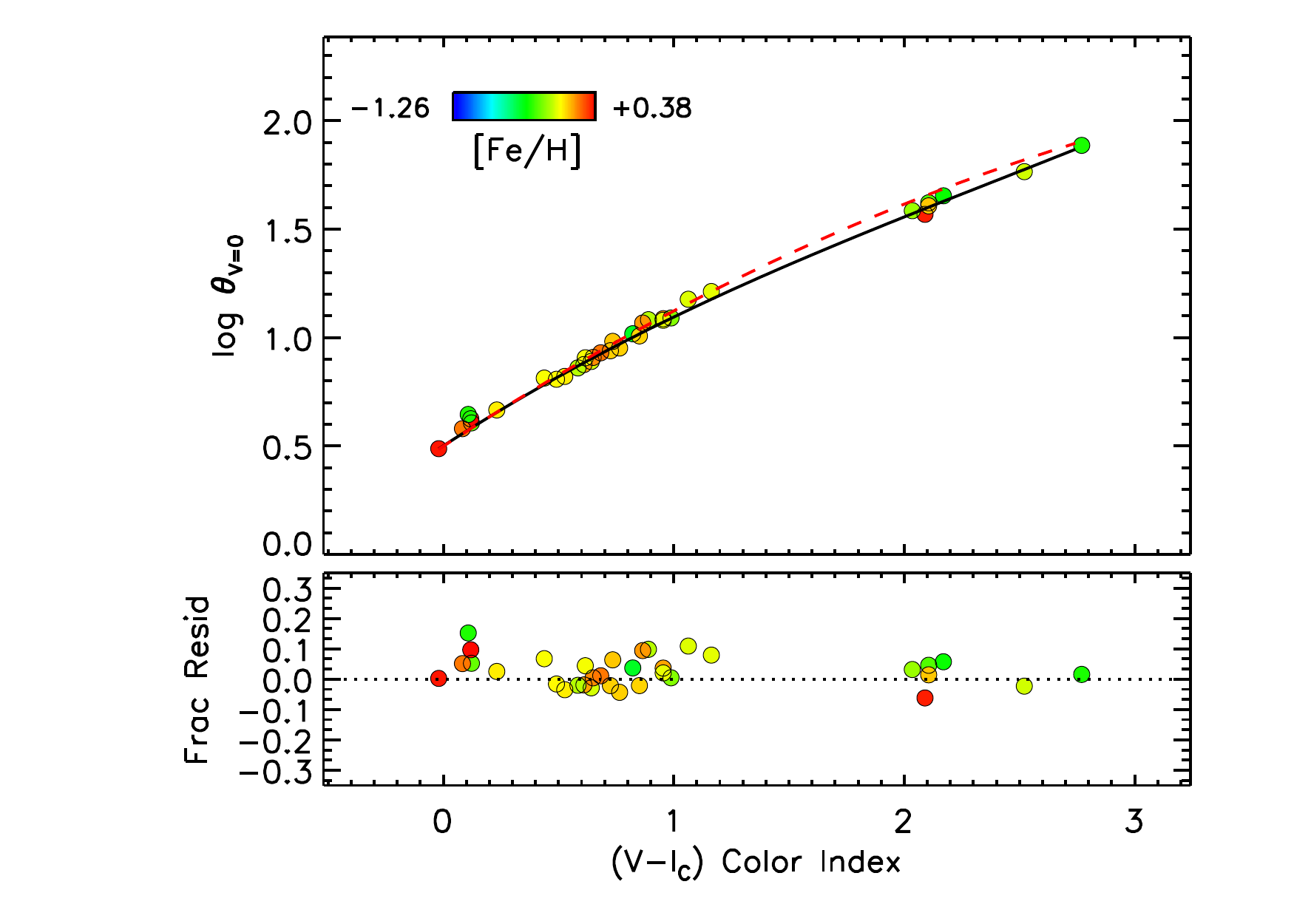}
\caption{Surface brightness relation for $V-I_c$ color. log$\theta_{V=0}$ corresponds to the angular diameter a star would have at a distance at which its $V=0$. The black curve illustrates the best fit to the data (4th order polynomial), and the colors of the data points represent stellar metalliticy values according to the scale on the top of the panel. The red, dashed line displays a previously published fit \citep{ker08}. Figure adapted from \citet{boy14}. See \S \ref{chapter:future:sec:methods} for more details.}
\label{fig:theta_vs_V-Ic}
\end{figure}

One example of such relations, between broad-band colors and effective temperatures for main sequence stars, is presented in \citet{boy13}, and we show in Fig. \ref{fig:Temp_vs_V-Ic} their results for $V-I_c$ color\footnote{$I_c$ represents Cousins $I$.} vs stellar effective temperature. The black line corresponds to the best fit, which is a 3rd order polynomial based on 34 stars fulfilling various criteria. The scatter of data around the fit is approximately 3\%. The different colors of the data points correspond to different metallicity values -- the scale is the same as shown in Fig. \ref{fig:theta_vs_V-Ic}. Similar fits are given in \citet{boy13} for combinations of 43 different broad-band photometric color indices. 

Another set of such relations is published in \citet{boy14} that link 48 different broad-band color indices to angular diameters of main sequence stars -- the so-called surface brightness relations \citep[SBRs; based on][]{wes69}. Fig. \ref{fig:theta_vs_V-Ic} shows their relation between $V-I_c$ color and the logarithm of the zero-magnitude angular diameter, $\theta_{V=0}$, which corresponds to the angular diameter that a star would have if it were at a distance at which its apparent magnitude $V$ is zero. The black curve illustrates the best fit to the data (4th order polynomial), and the colors of the data points represent stellar metalliticy values according to the scale on the top of the panel. The red, dashed line displays a previously published fit for SBRs.

While the SBRs in \citet{boy14} are restricted to main sequence stars, work is currently in progress that updates the \citet{boy14} equations with more data and combines the SBRs to encompass both main sequence and giant stars \citep{ada16}. 


%
%
%
\chapter{Summary and Conclusion}
\label{chapter:conclusion} 


\noindent
{\it In order to understand the exoplanet, you need to understand its parent star.}
\\
\\
Almost a decade ago, we used this statement as motivation to approach the challenge of studying exoplanet host stars by using as few assumptions as possible. The use of interferometry to determine their diameters meets this challenge despite its being limited to nearby and bright stars and the fact that it is intrinsically a complicated method (\S \ref{chapter:parameters:sec:radius}). Imperfect accuracy and/or precision of the method due to calibration challenges as function of atmospheric conditions, consequent characterization of associated uncertainties, and random errors in limb-darkening corrections certainly appear to outweigh the caveats of any other method of determining stellar diameters, at least for single stars (\S\ref{chapter:parameters:sec:why}). 

The determination of stellar effective temperature via the measurement of bolometric flux is the second component in our ``as directly as possible" approach (\S \ref{chapter:parameters:sec:temp}). Here, reduction in accuracy and/or precision in the result based on SED fitting are due to unknown systematics in the literature photometry data, the manual choice of spectral template in the absence of spectrophotometry, and the lack of grid and wavelength coverage of available spectral templates. Again, these potential pitfalls seem to outweigh their counterparts frequently present in less direct methods (\S \ref{chapter:parameters:sec:why}). The increasing availability of flux-calibrated spectrophotometry data is very helpful in reducing any potential systematic errors in the SED fitting due to undetected spectral features. 

Stellar radius and effective temperature are the most fundamental, directly determinable physical parameters from which a number of other quantities can be derived. The knowledge of the stellar energy output per unit time provides a full characterization of the radiation environment in which exoplanets reside. Any definition of system HZ (\S \ref{chapter:parameters:sec:hz}) will require this radiation environment as an input, irrespective of other assumptions such as the existence of a planetary atmosphere or similar. 

The knowledge of directly determined stellar astrophysical parameters allows the calculation of just about all system parameters if the exoplanet is transiting and there are literature RV and/or photometry data of sufficient precision -- see Sections \ref{chapter:systems:sec:gj436}, \ref{chapter:systems:sec:hd209458}, and \ref{chapter:systems:sec:hd189733} for examples. Even for stars too faint and/or too distant, relations calibrated with the methods described above allow for semi-empirical determination of stellar parameters, like the ones relating broad-band photometric colors to stellar effective temperatures or angular diameters  (Section \ref{chapter:future:sec:methods}; Figures \ref{fig:Temp_vs_V-Ic} and \ref{fig:theta_vs_V-Ic}) or the creation of isochrones based on stellar models (Section \ref{chapter:parameters:sec:why}). 

We provide a comprehensive overview of interferometrically determined, high-precision stellar diameters in Table \ref{tab:systems:dwarfs} in Section \ref{chapter:systems:sec:status}, current as of November 2016. This table contains around 300 stars, many with multiple measurements of their diameters. We use angular diameter and bolometric flux measurements from the respective publications and trigonometric parallax values from Hipparcos to follow our approach of being as direct as possible. Thence we uniformly calculate stellar physical radii, effective temperatures, and luminosities unless otherwise indicated in the Table, using weighted averages for multiple measurements. Of the roughly 300 stars in the table, around 150 are main-sequence stars and around 150 are subgiants or giants. Approximately 60 of them have known exoplanets in orbit. The graphical visualization of Table \ref{tab:systems:dwarfs} is shown in Figure \ref{fig:HRD_EHS_blue} where (1) the diameter of each data point is representative of the logarithm of the corresponding stellar radius, and (2) known exoplanet hosting stars are colored blue. 

Section \ref{chapter:systems:sec:individual} contains a number of individually discussed exoplanet host systems that are interesting from a historical or astrophysical perspective. These systems are somewhat arbitrarily chosen but for obvious reasons include (1) nearby stars since they were or can be studied with other methods to gain comprehensive insight into the system as a whole, (2) late-type dwarfs since they not only provide some much needed constraints to stellar models or semi-empirical methods but also shed light onto the nature of  architectures of late dwarfs and their exoplanets, (3) transiting planets since they allow a full system characterization when combined with other literature data, (4) multiplanet systems and/or systems with one or more planets in the HZ, and (5) exoplanet hosts that were historically significant in the field of exoplanet astronomy. 

Our work with the CHARA Array over the last decade has contributed the majority of directly determined, high-precision diameters and effective temperatures of main-sequence and particularly low-mass stars with and without exoplanets. With ongoing improvements to the system in terms of resolution and sensitivity, the number of available targets is continually increasing to allow for this work to continue far into the future. Current and upcoming space missions such as K2, TESS, PLATO, CHEOPS, etc. provide and will provide a variety of data products relating to stellar and exoplanet science at unprecedented levels of precision. Knowledge of stellar astrophysical parameters that is as accurate as possible will be paramount to the correct interpretation of these data products. 




\backmatter
%
%

\Extrachap{Glossary}

The following is a glossary of some of the terms used in this book. For additional definitions, we recommend the use of \url{https://wikipedia.org}, particularly their astronomy glossary page at \url{https://en.wikipedia.org/wiki/Glossary_of_astronomy}. 


\runinhead{Apastron} The point in the orbit of an extrasolar planet when it is the furthest away from its parent star. 

\runinhead{Asteroseismology} Study of stellar interiors and astrophysical parameters based on oscillations on their surfaces. See \S\ref{chapter:parameters:sec:radius}.

\runinhead{Baseline} The distance between two telescopes in an interferometric array, sometimes measured in units of the operational wavelength. See \S\ref{chapter:parameters:sec:radius}.

\runinhead{Bolometric Flux} The amount of energy received per second from a star, integrated across the full range of wavelengths. We determine this quantity by means of SED fitting. See \S\ref{chapter:parameters:sec:temp}.

\runinhead{CHARA} CHARA stands for Center for High Angular Resolution Astronomy and refers to the interferometric array on Mount Wilson in CA, owned and operated by Georgia State University, with which the majority of the work described in this book was performed. 

\runinhead{Coherent, Coherence} Electromagnetic radiation is said to be coherent if the waves maintain a constant phase relationship. This enables measuring the phase difference of a wavefront at two or more different locations on the ground, which constitutes the basic data required for interferometry. See \S\ref{chapter:parameters:sec:radius}.

\runinhead{Convective Mixing Length Parameter} The convective mixing length parameter, $\alpha_{MLT}$, refers to the distance that a parcel or blob of gas of a certain temperature would travel before it dissipates its thermal energy and adjusts its temperature to its surroundings. Thus, it gives a sense of how quickly a stellar atmosphere is mixed by convection. See \ref{chapter:systems:sec:hd189733}.

\runinhead{Effective Temperature} Stellar effective temperature, \teff, is defined as the surface temperature of a black body that emits as much energy per second as the star to which the effective temperature pertains. Thus, it provides a uniform measure of stellar temperature. See \S\ref{chapter:parameters:sec:temp}.

\runinhead{Extrasolar Planet, Exoplanet} Defined here as a planet that orbits a star or multiple stars other than the sun, though the term could also include free-floating planets, i.e., planets that do not orbit any stars. 

\runinhead{Habitable Planet} Defined here as a planet on which life can exist. Based on this definition, this currently only includes Earth. See \S\ref{chapter:parameters:sec:hz}.

\runinhead{Habitable Zone, HZ} The range of distances from its parent star(s) at which an exoplanet with a surface may harbor liquid water. Additional assumptions, such as the amount of greenhouse gases in the planetary atmosphere, go into the calculations of HZs. See \S\ref{chapter:parameters:sec:hz}.

\runinhead{Interferometric Fringe} The technique of interferometry is based on the detection of interference of coherent light to measure the brightness distribution of an object on the sky. Interference is visualized in interferometric fringes. See \S\ref{chapter:parameters:sec:radius} and Figure \ref{fig:fringe}.

\runinhead{Interferometer, Interferometric Array} The system that performs the technique of interferometry. See detailed description in \S\ref{chapter:parameters:sec:radius}.

\runinhead{Interferometry} The method of using multiple telescopes to achieve very high angular resolution. This technique is described in detail in \S\ref{chapter:parameters:sec:radius}.

\runinhead{Limb Darkening} Limb darkening refers to the effect where a stellar disk appears brighter at the center than at the limb. The reason for this effect is that one sees deeper and thus hotter layers of the stellar atmosphere at the center of the disk than at the limb. See \S\ref{chapter:parameters:sec:radius} and Figure \ref{fig:mercury}. 

\runinhead{Luminosity} Stellar luminosity refers to the amount of energy a star emits per unit time. 

\runinhead{Optical Delay Lines} These are elements of an interferometer to detect interferometric fringes. See \S\ref{chapter:parameters:sec:radius} for much more detail. See Figures \ref{fig:ople1} and \ref{fig:ople2} for what optical delay lines look like, and see Figure \ref{fig:schematics} for the function of optical delay lines in an interferometric array. 

\runinhead{Optical Path, Optical Path Length} The optical path is the path a photon takes from the star through all of the elements of the optical system to the detector. One of the principal engineering challenges involved in the operation of an interferometric array is the adjustment of the optical path lengths for all operating telescopes to exacly the same value via the use of optical delay lines. See \S\ref{chapter:parameters:sec:radius} and Figure \ref{fig:schematics}.

\runinhead{Periastron} The point in the orbit of an extrasolar planet when it is the closest to its parent star. 

\runinhead{Spectral Energy Distribution, SED, SED Fitting} The distribution of energy emitted by a star as a function of wavelength. The integral over wavelength of a stellar SED corresponds to the bolometric flux. To determine the zeropoint of the SED, we typically use literature broadband photometry, a process referred to as SED fitting. See \S\ref{chapter:parameters:sec:temp} for much more details on SEDs and SED fitting, and see Figures \ref{fig:hd189733_sed} and \ref{fig:gj614_sed} for examples. 

\runinhead{Spectrophotometry} In astronomy, spectrophotometry essentially refers to flux-calibrated spectroscopy. That is, the energy received at every wavelength is normalized such that relative energy levels are preserved -- in contrast to broad-band photometry where the energy received over a range of wavelengths is integrated across the filter bandwidth. As such, spectrophotometry provides a much less coarse approach to determining stellar energy distribution. See \S\ref{chapter:parameters:sec:temp} and Figure \ref{fig:hd189733_sed}.

\runinhead{Stellar Angular Diameter} Angular diameter refers to the apparent size of a star on the sky, as opposed to physical diameter, which is expressed in units of length. CHARA routinely measures stellar angular diameters of fractions of a milliarcsecond with 1--3\% precision. The size of a soccer ball as seen on the surface of the moon approximately corresponds to one milliarcsecond. See \S\ref{chapter:parameters:sec:radius}. 

\runinhead{Transiting Planet, Transiting Exoplanet} A planet is said to transit its parent star(s) if it partially occults the stellar surface as seen from an observer. Venus and Mercury periodically transit the surface of the sun as shown in Figure \ref{fig:mercury}. Transiting exoplanets block minute amounts of the light received from their parent stars. This flux decrement determines the relative sizes of planet and parent star and provides fundamental insights in exoplanet studies. 

\runinhead{van Cittert-Zernike Theorem} The van Cittert-Zernike Theorem represents the basis on which astronomical interferomety is founded. It essentially states that interferometric visibility is related to the brightness distribution of an object on the sky. See \S\ref{chapter:parameters:sec:radius}. 

\runinhead{Visibility, Visibility Function} Interferometric visibility represents the degree of coherence of light received at two or more telescopes in an interferometric array. It corresponds to the interferometric data produced when studying angular sizes. It is a function of the angular size of the object, the operational wavelength, and the projected length of the distance between the telescopes. The visibility function corresponds to the dependence of the visibility upon baseline. Its shape depends on the topology, i.e., brightness distribution, of the object on the sky. For uniform disk profiles, the visibility function looks like what is shown in Figure \ref{fig:55Cnc_vis}. See \S\ref{chapter:parameters:sec:radius} for much more detail. 



\printindex


\bibliographystyle{mn2e}            
\bibliography{mn-jour,paper,charm} 


\end{document}